\documentclass[
reprint,
superscriptaddress,
preprintnumbers,
nofootinbib,
 amsmath,amssymb,
 aps,
]{revtex4-2}

\usepackage[english]{babel}
\usepackage{xcolor}
\usepackage[linktocpage, colorlinks, citecolor=blue, linkcolor=blue, urlcolor=blue]{hyperref} 
\usepackage{enumerate}  
\usepackage{float}
\usepackage{calc}
\usepackage{dutchcal}
\usepackage{revsymb}
\usepackage{comment}
\usepackage{upgreek}
\usepackage{IEEEtrantools}
\usepackage{Commands}
\usepackage{soul} 
\usepackage[normalem]{ulem}
\usepackage{amsthm}

\DeclareSymbolFont{cmbrightop}{OT1}{cmbr}{m}{n}
\DeclareMathSymbol{\sfPsi}{\mathalpha}{cmbrightop}{9}

\usepackage{amsfonts, amssymb, amsthm, mathtools, mathrsfs}
\usepackage{physics}
\usepackage{bm}

\usepackage[caption=false]{subfig}
\usepackage{graphicx}
\usepackage{dcolumn}
\usepackage{booktabs}

\newcommand{\orcid}[1]{\href{https://orcid.org/#1}{\includegraphics[width=8pt]{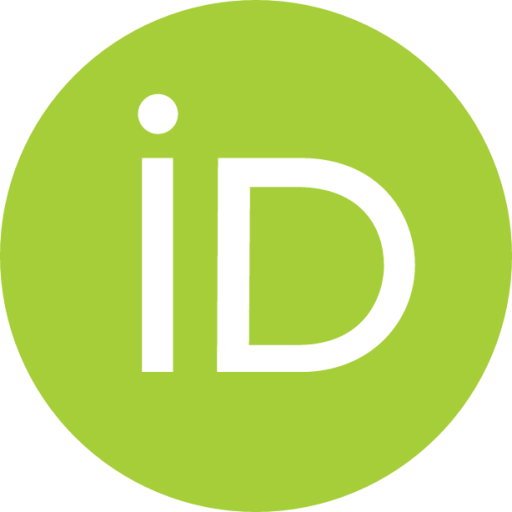}}}

\begin{document}
\preprint{CERN-TH-2026-104}

\title{Scalar memory from compact binary coalescences}
\author{Jann Zosso \orcid{0000-0002-2671-7531}}
\email{jann.zosso@nbi.ku.dk}
\affiliation{Center of Gravity, Niels Bohr Institute, Blegdamsvej 17, 2100 Copenhagen, Denmark}
\author{Silvia Gasparotto~\orcid{0000-0001-7586-1786}}
\email{silvia.gasparotto@cern.ch}
\affiliation{CERN, Theoretical Physics Department, Esplanade des Particules 1, Geneva 1211, Switzerland}
\author{Llibert Aresté Saló~\orcid{0000-0002-3812-8523}}
\affiliation{Instituut voor Theoretische Fysica, KU Leuven. Celestijnenlaan 200D, B-3001 Leuven, Belgium}
\affiliation{Leuven Gravity Institute, KU Leuven. Celestijnenlaan 200D, B-3001 Leuven, Belgium}
\author{Daniela D. Doneva~\orcid{0000-0001-6519-000X}}
\affiliation{Departamento de Astronom\'ia y Astrof\'isica, Universitat de Val\`encia,
Av. Vicent Andrés Estellés, 19, 46100, Burjassot (Val\`encia), Spain}
\affiliation{Theoretical Astrophysics, Eberhard Karls University of T\"ubingen, 72076 T\"ubingen, Germany}
\author{Stoytcho S. Yazadjiev~\orcid{0000-0002-1280-9013}  }
\affiliation{Department of Theoretical Physics, Sofia University ``St. Kliment Ohridski", Sofia 1164, Bulgaria}
\affiliation{Institute of Mathematics and Informatics, Bulgarian Academy of Sciences, Acad. G. Bonchev St. 8, Sofia 1113, Bulgaria}
\date{\today}

\begin{abstract}
Gravitational memory provides a distinctive low-frequency probe of gravity, but explicit merger studies beyond general relativity remain limited. In this letter,
we investigate memory from binary black hole mergers in Ricci-coupled scalar-Gauss-Bonnet gravity, a natural extension of scalar-Gauss-Bonnet theory that admits an additional scalar breathing polarization. Based on numerical-relativity waveforms of binary black hole coalescences, we show that the change in the scalar charge of the system across merger generates a significant scalar-memory contribution. For a GW150914-like system, this effect modifies the memory signal in a gravitational-wave detector on the same observable timescale and by an amount comparable to the pure scalar-Gauss-Bonnet correction to tensor memory. Thus, it can substantially enhance the total deviation from the general-relativity prediction over a broad range of source and detector configurations. We argue that this identifies a general mechanism: whenever a compact-binary merger changes the asymptotic charge of an additional gravitational field, and that field sources an observable extra polarization, the resulting memory can provide a leading low-frequency signature of new gravitational physics.
\end{abstract}

\maketitle



\section{Introduction}

With the rapid development of gravitational-wave astronomy, compact-binary coalescences have become precision probes of strong-field gravity~\cite{LIGOScientific:2021sio,LIGOScientific:2025slb,LIGOScientific:2026qni}.
Future detectors, including Einstein Telescope~\cite{ET:2025xjr}, Cosmic Explorer~\cite{Reitze:2019iox}, LISA~\cite{LISA:2024hlh} and TianQin~\cite{TianQin:2020hid}, will considerably extend both the number of detectable events and the precision with which new signatures of gravitational physics can be tested. Among such signatures, the gravitational memory effect is especially interesting. Memory corresponds to a permanent displacement of freely falling test masses after the passage of a burst of gravitational radiation~\cite{Zeldovich:1974gvh,Braginsky:1985vlg,Braginsky:1987kwo,PhysRevD.45.520,Christodoulou:1991cr,Blanchet:1992br,Wiseman:1991ss,Favata:2009ii}. In general relativity, the dominant nonlinear displacement memory is sourced by radiative energy flux reaching null infinity and is closely related to asymptotic Bondi-Metzner-Sachs (BMS) symmetries and soft theorems~\cite{Bondi:1962px,Sachs:1962wk,Weinberg_PhysRev.140.B516,Strominger:2017zoo,Strominger:2013jfa,He:2014laa,Strominger:2014pwa,Pasterski:2015tva}. Although its low-frequency character makes detection challenging~\cite{Hubner:2019sly,Hubner:2021amk,Grant:2022bla,Gasparotto:2023fcg,Inchauspe:2024ibs,Cogez:2026frh,Zosso:2026czc}, the prospects for a first measurement in next-generation detectors are encouraging. This raises a natural question: how can gravitational memory be used to test gravity beyond general relativity?

A general theoretical framework for memory beyond GR has recently been developed for Horndeski gravity and more general scalar-vector-tensor effective field theories~\cite{Heisenberg:2023prj,Heisenberg:2024cjk,Zosso:2024xgy,Zosso:2025ffy,Heisenberg:2025tfh,Heisenberg:2025roe,Maibach:2026wpz}. Together with previous works on specific scalar-tensor theories~\cite{Lang:2013fna,Lang:2014osa,Du:2016hww,Sennett:2016klh,Tahura:2020vsa,Tahura:2021hbk,Hou:2020tnd,Hou:2020wbo,Koyama:2020vfc,Seraj:2021qja,Bernard:2022noq,Trestini:2024zpi,Ma:2024bed,Tahura:2025ebb}, these results show that additional radiative degrees of freedom (DOFs) can modify the GR expectation both through direct corrections to tensor memory and by admitting new types of memory in the additional polarization states of the asymptotic radiation, including \emph{scalar memory} \cite{Du:2016hww,Tahura:2020vsa,Tahura:2021hbk,Hou:2020tnd,Hou:2020wbo,Koyama:2020vfc,Seraj:2021qja,Bernard:2022noq,Heisenberg:2023prj,Heisenberg:2024cjk,Tahura:2025ebb}. Concrete mechanisms for generating scalar memory have been proposed, for instance in Brans-Dicke-type theories from the collapse of compact objects to a black hole, where the no-hair property of the final black hole ties the scalar offset to the disappearance of the progenitor scalar charge~\cite{Du:2016hww,Koyama:2020vfc}. However, explicit inspiral--merger--ringdown studies of memory beyond GR, which are especially relevant for applications to GW data, remain scarce.

In this context, our recent work~\cite{Gasparotto:2026bru} employed numerical-relativity simulations within scalar-Gauss-Bonnet (sGB) gravity to study the corresponding nonlinear memory from binary black hole (BBH) mergers. This theory is motivated both from a bottom-up perspective, as the leading ghost-free scalar-curvature correction in a parity-even derivative expansion of gravity~\cite{Weinberg:2008hq,Simon:1990PhysRevD41,Yunes:2013dva,Zosso:2024xgy}, and from a top-down perspective, where sGB couplings naturally arise in the low-energy effective actions of heterotic and bosonic string compactifications~\cite{Zwiebach:1985uq,Gross:1986iv,Boulware:1986dr,Metsaev:1987zx,Torii:1996yi,Moura:2006pz,Nojiri:2017ncd,Ortega:2024prv}. Yet, if this high-energy motivation is taken seriously, the usual pure sGB formulation is incomplete, since in the physical Jordan frame the dilaton multiplies not only the Gauss--Bonnet term but also the Ricci scalar, leading naturally to a Ricci-coupled scalar-Gauss-Bonnet (RCsGB) theory~\cite{Metsaev:1987zx,Antoniou:2021zoy,Evstafyeva:2022rve,Doneva:2024ntw}. This additional coupling has an important phenomenological consequence: unlike pure sGB gravity, it makes the scalar degree of freedom directly observable in geodesic deviation through a breathing polarization.

The central contribution of this work is to show that this additional polarization opens a qualitatively new channel for observable memory from compact-binary coalescences (CBCs). In Ricci-coupled sGB gravity, the merger-induced relaxation of the scalar monopole produces a scalar memory contribution in the form of a permanent offset in the breathing polarization, which can significantly affect the total memory in the detector response. This identifies a new general mechanism: whenever a merger reorganizes the asymptotic charge of an additional gravitational field and that field sources an observable extra polarization, the resulting memory can play a leading role in the low-frequency detector response deviation from GR and provide valuable information on the presence of additional gravitational DOFs.

\section{Ricci-coupled scalar-Gauss-Bonnet gravity}\label{Sec:RCsGB}

\subsection{Physical Jordan frame}

We consider Ricci-coupled scalar-Gauss-Bonnet gravity in the physical \emph{Jordan frame}, defined in geometric units as \cite{Metsaev:1987zx,Moura:2006pz,Ortega:2024prv}
\begin{align}\label{eq:ActionsString}
    S=\frac{1}{16\pi}&\int d^4 x\sqrt{-g}\,f(\Phi)\bigg( R+\omega(\Phi)X-\lambda \mathcal G\bigg)\nonumber\\ &+S_\text{m}[g,\Psi_\text{m}]\,,
\end{align}
where $X\equiv - \frac{1}{2} \partial_\mu\Phi\partial^\mu\Phi$, $R$ is the Ricci scalar of the physical metric $g$ and $\Phi$ is a scalar field. $f(\Phi)$ and $\omega(\Phi)$ are assumed to be sufficiently well behaved smooth functions. The Gauss-Bonnet curvature scalar is defined as 
\begin{equation}
    \mathcal{G}\equiv 
    R^{\mu\nu\rho\sigma}R_{\mu\nu\rho\sigma}-4R^{\mu\nu}R_{\mu\nu}+R^2\,.
\end{equation}
Finally, $S_\text{m}[g,\Psi_\text{m}]$ is the matter action depending on generic matter fields $\Psi_m$ that are universally and minimally coupled to the metric, rendering the theory a so-called \emph{metric theory} of gravity \cite{papantonopoulos2014modifications,Will:2018bme,Zosso:2024xgy}.

The equations of motion of this theory remain at second order per field and therefore the theory is included in the ghost-free Horndeski class of scalar-tensor theories \cite{Horndeski:1974wa,Nicolis:2008in,Deffayet:2009wt,Deffayet:2009mn,Heisenberg:2018vsk,Kobayashi:2019hrl}. It admits a well-posed formulation of its full equations of motion, provided the evolution remains within the regime of validity of the effective field theory so that the higher-curvature operators act as subleading corrections to the principal two-derivative dynamics \cite{Kovacs:2020ywu,Kovacs:2020pns,Ripley:2022cdh}. 
In Appendix~\ref{App:Horndeski} we provide the explicit embedding of Eq.~\eqref{eq:ActionsString} into the Horndeski framework.

As discussed in the introduction, the Ricci-coupled action in Eq.~\eqref{eq:ActionsString} with a non-minimal coupling of the form
\begin{equation}\label{eq:nonminimalcoupling1}
    f_\alpha(\Phi)=e^{-\alpha\Phi}\,
\end{equation}
is inspired from its appearance within the Jordan frame (or \emph{string frame}) at leading order in a general $4d$ compactification of both heterotic and bosonic string theories.\footnote{Although also the parity-violating Chern-Simons term \cite{Jackiw:2003pm,Alexander:2009tp} together with the axion, as well as a higher derivative dilaton terms appear \cite{Metsaev:1987zx,Ortega:2024prv}, the sGB term is the only viable combination to form a ghost-free and hence exact theory of gravity \cite{Zwiebach:1985uq,Moura:2006pz}.} 
Beyond the high-energy physics motivation, we choose to consider Ricci-coupled sGB gravity as a minimal example theory that combines the existence of well-defined scalar-charged black hole solutions with the admission of additional gravitational polarizations within the gravitational radiation.

\subsection{Einstein-frame formulation}\label{sSec:EinsteinFrame}

For vacuum binary-black-hole dynamics it is useful to remove the non-minimal Ricci coupling in Eq.~\eqref{eq:ActionsString} by a field-dependent Weyl rescaling to the Einstein frame
\begin{equation}\label{eq:WeylRescaling}
    \tilde g_{\mu\nu}=f(\Phi)g_{\mu\nu}\, .
\end{equation}
Specializing for concreteness to the string-motivated exponential coupling in Eq.~\eqref{eq:nonminimalcoupling1} and working within a small-coupling and small-scalar-field regime of the EFT expansion
\begin{equation}\label{eq:nonminimalcouplingExp}
    f_\alpha(\Phi)=e^{-\alpha\Phi}
    \simeq 1-\alpha\Phi\, ,
\end{equation}
the leading-order Einstein-frame action becomes
\begin{align}\label{eq:actionEinstein}
    S_E^\alpha
    =&\, \frac{1}{16\pi}
    \int d^4x \sqrt{-\tilde g}
    \left[
        \tilde R
        +X
        +\lambda_\alpha\Phi\tilde{\mathcal G}
    \right]
    \nonumber\\
    &+ S_\text{m}[(1+\alpha\Phi)\tilde g_{\mu\nu},\Psi_\text{m}]\,,
\end{align}
where
\begin{equation}
    \lambda_\alpha\equiv \alpha\lambda\, .
\end{equation}
For simplicity, we have also chosen here a canonically normalized Einstein-frame scalar. For a more general treatment of the theory and the explicit Weyl transformation we refer to Appendices~\ref{App:Horndeski} and \ref{App:Weyl details}.

In the Einstein-frame, the vacuum gravitational sector of Eq.~\eqref{eq:actionEinstein} may thus be reduced to a standard shift-symmetric sGB theory. Note that \(\lambda_\alpha/M^2\) for some characteristic mass scale $M$ is the relevant vacuum expansion parameter both in the Einstein and the Jordan frame. On the other hand, it is imperative to recall that under the Weyl rescaling in Eq.~\eqref{eq:WeylRescaling} the universal coupling property of matter is lost. The Einstein-frame metric \(\tilde g_{\mu\nu}\) therefore cannot be interpreted as describing an objective \emph{spacetime} that determines freely falling frames and locally recovers all Minkowskian non-gravitational physics as required by the Einstein equivalence principle~\cite{Will:2018bme,Zosso:2024xgy}. Gravitational observables must instead be defined with respect to the physical Jordan-frame metric \(g_{\mu\nu}\). This is particularly important for gravitational-wave detection, where the observable response is the geodesic deviation of freely falling test masses.

\subsection{Gravitational radiation and polarizations}\label{sSecRadiationAndPolarizations}

Let us now consider an asymptotically flat background spacetime in asymptotic source-centered Minkowski coordinates \(\{t,r,\Omega=(\theta,\phi)\}\), which we denote by \(\{\eta_{\mu\nu},\varphi_0\}\), where \(\varphi_0\) is a constant background value of the scalar field \(\Phi\). Perturbing a metric theory around this background up to \(\mathcal{O}(1/r)\), the leading-order geodesic deviation in terms of asymptotic retarded time \(u\equiv t-r\) can be written as \cite{misner_gravitation_1973,Flanagan:2005yc,maggiore2008gravitational,carroll2019spacetime,Heisenberg:2024cjk}
\begin{align}\label{eq:integratedGeodesicDeviation}
    \Delta s_i (u) = \frac{1}{2} \Delta P_{ij}(u) s^j (u_0) \,,
\end{align}
with \(\Delta s_i (u)\equiv s_i (u)-s_i (u_0)\). The response matrix can be expanded into at most six orthogonal polarization states \(\Lambda\), depending on the dynamical DOFs of the theory,
\begin{equation}\label{eq:PolarizationMatrixDef}
    P_{ij}=P_\Lambda\,e^\Lambda_{ij}\,,
\end{equation}
as dictated by its relation to the leading-order electric-parity part of the Riemann tensor of the physical metric \(g\) in the Jordan frame,
\begin{equation}
    R_{0i0j} = -\frac{1}{2} \ddot P_{ij} \,.
\end{equation}

Concretely, for the Ricci-coupled sGB theory considered in this work with leading-order massless tensor $h_+$, $h_\times$ and scalar $\varphi_1$ DOFs, the geodesic deviation is characterized by the response matrix [Appendix~\ref{App:Horndeski}]
\begin{align}\label{eq:Pijexpand}
P_{ij}
=
e^+_{ij}\,h_+
+
e^\times_{ij}\,h_\times
+
\alpha\,e^b_{ij}\,\varphi_1\, .
\end{align}
The three associated polarization tensors are provided explicitly in Eq.~\eqref{eq:PolTensors} of Appendix~\ref{sApp:RadiationBasis}. Hence, the additional non-minimal coupling to the Ricci scalar implies the existence of a transverse \textit{scalar breathing polarization} in the theory. That is, on top of the two propagating transverse-traceless tensor degrees of freedom $h_+$ and $h_\times$, also the propagating scalar degree of freedom $\varphi_1$ leaves its imprints in the leading geodesic deviation and can directly be observed in gravitational wave experiments. In contrast, in standard sGB gravity only the tensor radiation is directly accessible in GW data, while the scalar field may affect the radiation only indirectly through the vacuum dynamics, see the discussion in~\cite{Gasparotto:2026bru}.

\subsection{Observational constraints}\label{sSec:observationalConstraints}

For RCsGB gravity with a coupling function as defined in Eq.~\eqref{eq:nonminimalcoupling1}, current phenomenological viability requires both the non-minimal Ricci coupling parameter $\alpha$ 
and the combined Gauss-Bonnet coupling $\lambda_\alpha$ to be small. In particular,
weak-field Solar-System tests impose the strongest bound on the canonically normalized coupling $\alpha$, which controls the direct sourcing of the scalar field by any matter energy-momentum. Concretely the Cassini Shapiro time-delay measurements impose \cite{Bertotti:2003rm,Will:2014kxa,Evstafyeva:2022rve}
\begin{align}
    \alpha \lesssim \sqrt{10^{-5}} \approx 3\times 10^{-3} \, ,
\end{align}
while weaker bounds
$\alpha^2 \lesssim 10^{-4}$ are obtained from Very-Long-Baseline Interferometry (VLBI) measurements of solar light deflection \cite{PhysRevLett.92.121101}
and $\alpha^2 \lesssim 10^{-3}$ from the Viking time-delay experiment
\cite{Reasenberg:1979ey}.

In the strong-field sector, current gravitational-wave observations and binary pulsars constrain the corresponding sGB coupling to roughly \(\sqrt{|\lambda_\alpha|}\lesssim \mathcal{O}(1)\,{\rm km}\) \cite{Perkins:2021mhb,Lyu:2022gdr,Wang:2023wgv,Yordanov:2024lfk}.\footnote{The GW150914-like benchmark coupling \(\lambda_\alpha/m_2^2=0.1414\) employed in Section~\ref{Sec:MemoryBBH} below is thus slightly above the bound for the smallest known BH of $3.6M_{\odot}$, but remains sufficiently close to current strong-field bounds to serve as a useful proof-of-principle example, in particular accounting for the fact that these bounds rely on an uncertain PN theory extrapolation to merger \cite{Corman:2025wun}.} Requiring that sGB gravity remain within the validity of the effective field theory regime, and is thus hyperbolic, implies roughly 
\begin{align}
    \sqrt{|\lambda_{\alpha}|}\lesssim 3 \cdot\frac{m_2}{3.6M_{\odot}}\,\text{km}\,,
\end{align}
where $m_2$ is the mass of the smallest BH.


\section{Analytic memory formulas}

From an observational point of view, the gravitational memory effect can be defined as a permanent geodesic deviation within Eq.~\eqref{eq:integratedGeodesicDeviation},
\begin{equation}\label{eq:MemoryDef}
    \lim_{u\rightarrow\infty}\Delta s(u)\neq 0\,,
\end{equation}
implying a lasting change in the relative separation of freely-falling test masses after the passage of a transient gravitational radiation signal.
Therefore, the memory effect relies on a non-zero difference between the initial and final stages of the gravitational radiation within one of the components of the polarization matrix in Eq.~\eqref{eq:PolarizationMatrixDef},
\begin{equation}\label{eq:Defmemory}
    \lim_{u\rightarrow\infty} \Delta P_\Lambda(u)\neq 0\,.
\end{equation}
In this operational sense, gravitational memory is well defined in any metric theory of gravity, independent of the specific structure of the dynamical equations. What changes from theory to theory is the number and type of polarization channels that can contribute to the detector response and hence to the memory signal.

\subsection{Tensor memory}

First of all, RCsGB admits so-called \emph{tensor memory}, that is memory within the tensor polarizations $\Lambda=+,\times$. Indeed, tensor memory arises universally due to the emission of unbound energy-momentum from a localized source \cite{Christodoulou:1991cr}, a result that naturally generalizes to metric theories beyond GR \cite{Heisenberg:2023prj,Heisenberg:2024cjk}. In particular, the emission of tensor and scalar gravitational waves induces a low-frequency component in the asymptotic radiation which in our conventions for RCsGB gravity reads~\cite{Heisenberg:2023prj}
\begin{align}\label{eq:NonLinDispMemorysGB}
    \delta h_\lambda(u,r,\Omega)=\,\frac{1}{8\pi r} \int d^2\Omega'\,\frac{F(u,\Omega')\,e^{ij}_\lambda \,n'_in'_j}{1-\vec{n}'\cdot\vec{n}(\Omega)}\,,
\end{align}
where \(\vec n(\Omega)\) is the unit radial direction of propagation, $\vec{n}'\equiv \vec{n}(\Omega')$ and
\begin{equation}\label{eq:EperSolidAngleGR}
    F(u,\Omega')\equiv
    \int_{-\infty}^u d u' \, r'^2
    \left\langle |\dot{h}|^2+\dot{\varphi}_1^{\,2}\right\rangle .
\end{equation}
Here, we have defined
\begin{equation}
    h\equiv h_+-i h_\times\,,
\end{equation}
and introduced a spacetime averaging $\langle \ldots\rangle$ over the scales of variation of the emitted gravitational waves that singles out a non-oscillatory low frequency memory component \cite{Heisenberg:2023prj,Zosso:2025ffy,Zosso:2026czc}. Such an averaging operator naturally appears in the definition of a gauge-invariant localized GW energy-momentum \cite{Isaacson_PhysRev.166.1263,Isaacson_PhysRev.166.1272,misner_gravitation_1973,Flanagan:2005yc,maggiore2008gravitational}. 
Eq.~\eqref{eq:NonLinDispMemorysGB} represents the so-called tensor \emph{null memory} component within the radiation, which generally dominates the final memory offset within the geodesic deviation [Eq.~\eqref{eq:MemoryDef}]~\cite{Gasparotto:2026bru}.

\subsection{Scalar memory}

The presence of the additional scalar polarization within Eq.~\eqref{eq:Pijexpand} implies that RCsGB gravity also allows for \emph{scalar memory} defined as a permanent offset in the scalar-polarization component of the geodesic-deviation observable in Eq.~\eqref{eq:Defmemory}
\begin{equation}\label{eq:ScalarMemoryDef}
    \lim_{u\to\infty}\Delta P_b(u)
    =
    \alpha \lim_{u\to\infty}\Delta\varphi_1(u)
    \neq 0\,.
\end{equation}
While tensor memory is directly associated to the asymptotic  BMS spacetime symmetries, it was shown that scalar memory beyond the monopole scalar shift can also be associated to an asymptotic symmetry within a dual formulation of the scalar-tensor theory \cite{Seraj:2021qja,Maibach:2026wpz}.

Unlike tensor null memory, scalar breathing memory is, however, not reconstructed from a charge-flux integral analogous to Eq.~\eqref{eq:NonLinDispMemorysGB}~\cite{Heisenberg:2023prj,Zosso:2024xgy,Ma:2024bed}. Instead, in the case of non-conservation of scalar charge, scalar memory may directly arise from a change in the asymptotic \(\mathcal{O}(1/r)\) scalar monopole \cite{Du:2016hww}. Such a scalar-monopole contribution to memory is, however, more special and is not currently known to be associated with an analogous asymptotic symmetry. In the following, we show that such a monopole shift can nevertheless make a significant contribution to the observable memory in RCsGB compact binary coalescences.

\section{Gravitational memory of RCsGB binary black hole mergers}\label{Sec:MemoryBBH}

\subsection{Radiation from BBH mergers}
\subsubsection{Numerical relativity waveforms}

We use existing numerical-relativity waveforms from shift-symmetric sGB gravity as the dynamical input for our proof-of-principle study of memory in RCsGB gravity. Concretely, we employ the waveforms presented in Refs.~\cite{AresteSalo:2025sxc,Corman:2025wun} of BBH simulations performed with \texttt{GRFolres}~\cite{AresteSalo:2023hcp}, an extension of \texttt{GRChombo}~\cite{Andrade:2021rbd}.

This is justified because, as discussed in Section~\ref{sSec:EinsteinFrame}, the leading-order vacuum BBH dynamics of RCsGB coincide with those of standard shift-symmetric sGB gravity in the canonically normalized Einstein frame. The Ricci coupling instead changes how the scalar degree of freedom enters physical observables after transforming back to the Jordan frame, similar to previous studies of black hole mergers in Ricci-coupled sGB \cite{Evstafyeva:2022rve}. We therefore use the same simulated tensor and scalar waveforms as in the pure-sGB analysis of Ref.~\cite{Gasparotto:2026bru}, and isolate the RCsGB effect through the additional breathing term in the detector response in Eq.~\eqref{eq:Signal in Detector} below.

\begin{figure}[h]
    \centering
    \includegraphics[width=1\linewidth]{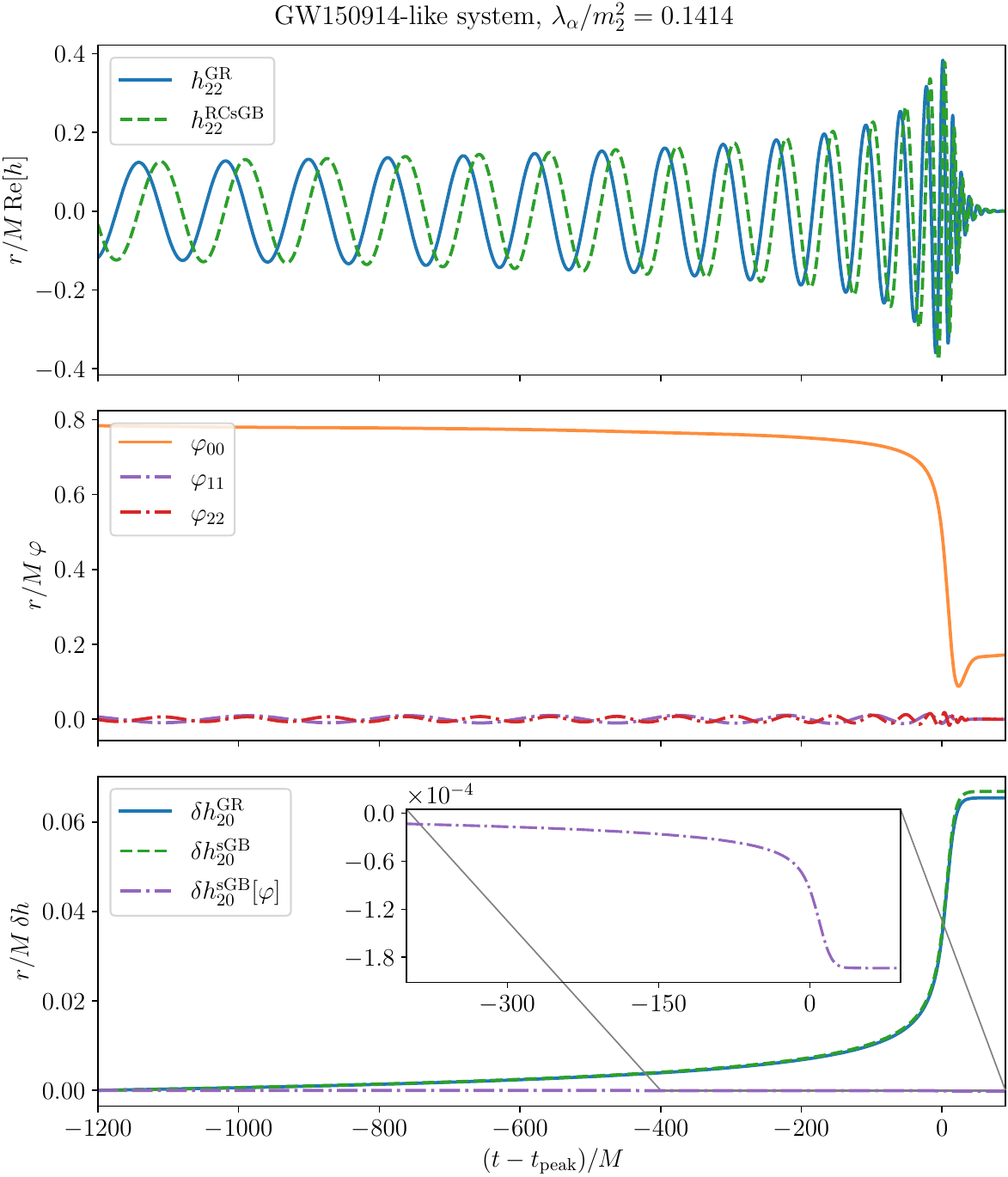}
    \caption{NR waveforms with \(\lambda_\alpha/m_2^2=0.1414\) and mass ratio \(q=1.221\) of a GW150914-like BBH, comparing GR and the vacuum dynamics of RCsGB. \emph{Top}: dominant \((2,2)\) tensor waveforms. \emph{Middle}: leading scalar multipoles. \emph{Bottom}: dominant tensor memory mode \(\delta h_{20}\), with the total sGB memory--corresponding to the \emph{tensor} memory in RCsGB--about \(2\%\) larger than in GR. The scalar-induced contribution \(\delta h^{\myst{sGB}}_{20}[\varphi]\), shown enlarged in the inset, is subleading and of opposite sign. The waveforms are aligned at the maximum of their dominant oscillatory signal.}
    \label{fig:Waveforms}
\end{figure}

\subsubsection{Concrete radiation events}

As a representative proof-of-principle example, we consider a GW150914-like BBH configuration from \cite{Corman:2025wun}, corresponding to a non-spinning near-equal-mass binary with mass ratio \(q=m_1/m_2=1.221\) at relatively large GB coupling \(\lambda_\alpha/m_2^2=0.1414\), where \(m_2\) denotes the ADM mass of the secondary black hole in the underlying Einstein-frame simulation.
As usual, we decompose the asymptotic tensor and scalar radiation in spin-weighted spherical harmonics (SWSH) in the reference frame of the source as
\begin{align}
h(u,r,\Omega) &\equiv h_+-ih_\times
   =\sum_{\ell,m} h_{\ell m}(u,r)\,{}_{-2}Y_{\ell m}(\Omega)\,,\\
\varphi_1(u,r,\Omega)
    &=\sum_{\ell,m}\varphi_{\ell m}(u,r)\,Y_{\ell m}(\Omega)\,.
\end{align}
\begin{figure*}
    \centering
    \includegraphics[width=1\textwidth]{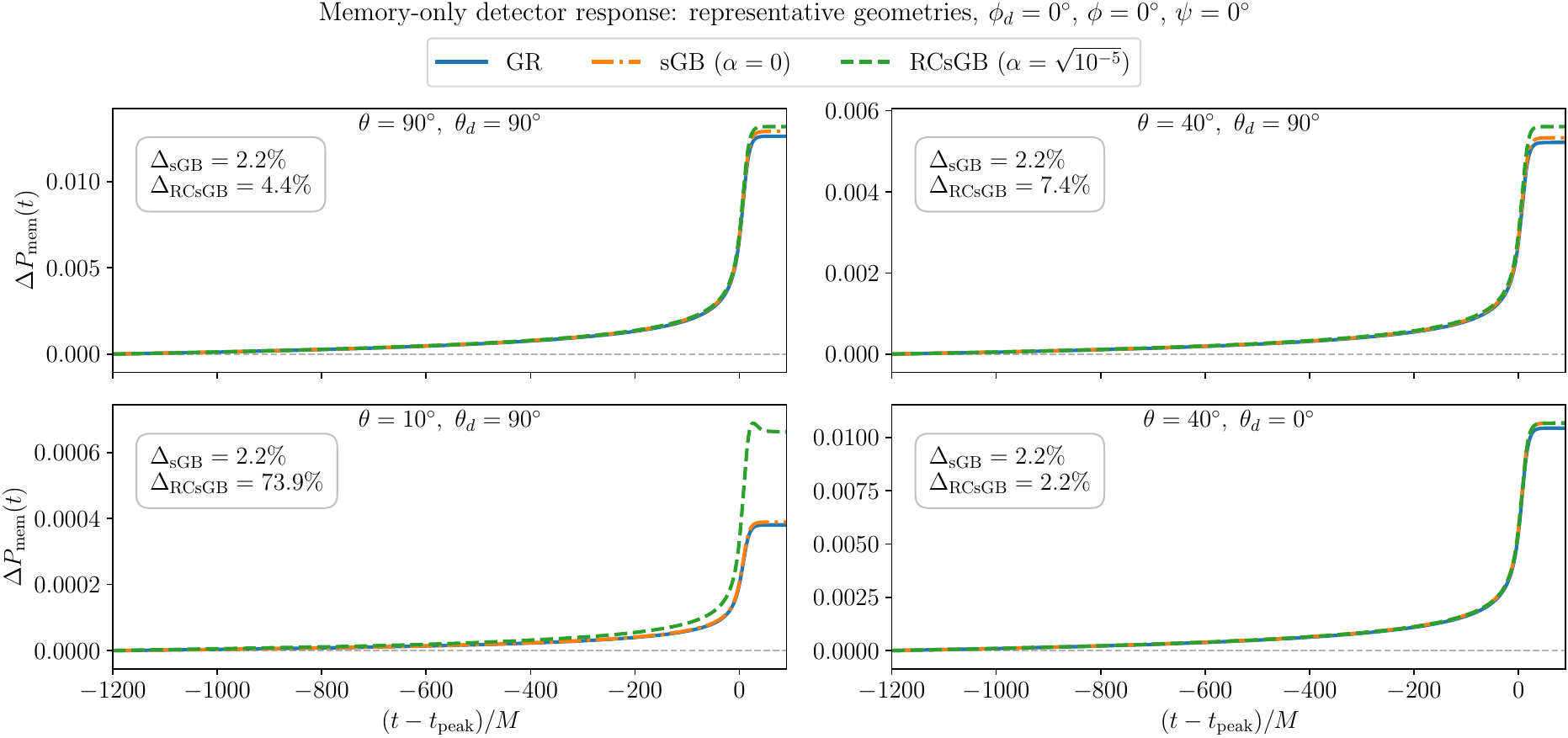}
    \caption{Memory-only detector response for four representative source-detector geometries. Shown are GR (solid blue), the pure-sGB response with vanishing scalar breathing coupling (dash-dotted orange), and RCsGB with \(\alpha=\sqrt{10^{-5}}\) (dashed green). The labels in each panel indicate the corresponding source inclination \(\theta\) and detector angle \(\theta_d\), while the boxes report the relative deviation of the total memory offset in pure sGB and RCsGB with respect to GR defined in Eq.~\eqref{eq:RelDevTotMemory}. \emph{Top left}: edge-on configuration with maximal tensor memory and maximal sensitivity to the scalar breathing mode. \emph{Top right}: decreasing the source inclination enhances the relative importance of the scalar-memory contribution. \emph{Bottom left}: for a more face-on system, the scalar memory dominates the beyond-GR correction, although the overall memory amplitude is reduced. \emph{Bottom right}: example of a detector orientation with vanishing sensitivity to scalar polarizations, shown for the same source inclination as in the top right panel.}
    \label{fig:DetectorResponse}
\end{figure*}

The top and middle panels in Fig.~\ref{fig:Waveforms} display the dominant \(h_{22}=h^*_{2,-2}\) tensor waveforms together with the leading scalar multipoles used in the analysis, comparing the GR and RCsGB radiation signals. The difference in the oscillatory tensor waveform remains small, at the sub-percent level around merger, although the RCsGB binary takes slightly longer to merge and reaches a mildly enhanced amplitude. In the scalar sector, the dominant contribution in terms of amplitude is the monopole \(\varphi_{00}\), while the \(\varphi_{11}\) and \(\varphi_{22}\) modes are suppressed by more than one order of magnitude. The prominence of the monopole reflects the net change in the asymptotic scalar charge carried by the system across merger.

The lower panel in Fig.~\ref{fig:Waveforms} shows the corresponding dominant tensor memory mode \(\delta h_{20}\) in GR and RCsGB, which corresponds to the total memory in pure sGB, together with the scalar-induced contribution to the tensor memory. Indeed, according to Eq.~\eqref{eq:NonLinDispMemorysGB}, both tensor and scalar GWs source a tensor memory contribution. In practice, the tensor memory mode \(\delta h_{20}\) is computed via Eq.~(30) of Ref.~\cite{Gasparotto:2026bru}. Consistent with Ref.~\cite{Gasparotto:2026bru}, the tensor memory in RCsGB is larger than in GR, with a relative enhancement of order a few percent. This difference is dominated by the modification of the emitted tensor GWs, while the scalar contribution to the tensor memory, which contributes with opposite sign, remains subleading. In other words, there is a clear hierarchy between the GB-induced modification of the nonlinear merger dynamics and the direct back-reaction from scalar energy flux. 

The total shift in tensor memory may already be observationally relevant. Ref.~\cite{Gasparotto:2026bru} found, through a detector-oriented mismatch analysis in which the GR template was optimized over the total mass, that for Einstein Telescope (ET) the inclusion of memory increases the GR-sGB mismatch by more than an order of magnitude, together with an increased effective memory deviation of order a few percent. Consistently, ET is expected to constrain the amplitude of displacement memory at the \(\sim2\%\) level~\cite{Goncharov:2023woe}, suggesting that sGB-induced memory shifts may be within reach of next-generation detectors.

In addition to the GW150914-like event, we use three unequal-mass simulation sets from \cite{AresteSalo:2025sxc}, already employed in Ref.~\cite{Gasparotto:2026bru}, with mass ratios \(q=1,2,3\), to probe the dependence on binary asymmetry. These runs are performed at fixed dimensionless coupling \(\lambda_\alpha/m_2^2=0.106\), and are compared to the corresponding GR evolutions with \(\lambda_\alpha=0\). For further details on the simulation setup, waveform alignment, extraction and robustness tests of the underlying sGB dataset, we refer to Refs.~\cite{AresteSalo:2025sxc,Gasparotto:2026bru}.

\subsection{Multi-polarization memory response}\label{sSec:DetectorResponse}

\subsubsection{Detector geometry}
To quantify the effect of the additional scalar polarization on memory, we consider the specific detector geometry of an equal-arm rectangular interferometer, which provides the basic building block of current and future ground-based detector networks.
In the short-timescale and long-wavelength approximation relevant for BBH memory, the associated \emph{detector response function}\footnote{For a detector whose arms subtend an angle $\chi$, the overall detector response is simply rescaled by a factor of $\sin\chi$ \cite{poisson2014gravity}.} is defined as
\begin{align}\label{eq:Signal in Detector}
P(t)=\,& F_+(\Omega_d)\,h_+(t,\Omega)+F_\times(\Omega_d)\,h_\times(t,\Omega)\nonumber\\
&+F_b(\Omega_d)\,\alpha \,\varphi_1(t,\Omega)\,,
\end{align}
with antenna patterns as a function of the detector-frame spherical angles $\Omega_d=\{\theta_d,\phi_d\}$ given by Eqs.~\eqref{eq:DetectorPatternApp}.

The antenna patterns also depend on the polarization angle $\psi$, which parameterizes the rotational freedom of the polarization basis in the plane transverse to the propagation direction. Without loss of generality, we set $\psi = 0$ throughout this work. The absolute values of these antenna pattern functions are also shown in Appendix~\ref{sApp:AntennaPattern} in Fig.~\ref{fig:antenna_patterns}.
In the following, we also set \(\phi_d=0\) when presenting explicit detector-response examples. This choice is sufficient to display the essential angular dependence, since for a network such as Einstein Telescope, the azimuthal dependence is effectively averaged over by the detector geometry. Moreover, without loss of generality, we choose \(\phi=0\) for the source azimuth angle.

\subsubsection{Memory in RCsGB gravity}

We now come to the central result of this work: in RCsGB gravity, the observable memory in the detector response receives not only the beyond-GR correction already present in pure sGB through the tensor sector, but in addition a scalar-memory contribution associated with the net change of scalar charge across merger. In the present example, this additional contribution enters at the same parametric order in the detector response as the tensor-memory modification and can therefore substantially enhance the total departure from the GR expectation.

For the numerical waveforms considered here, the monopole \(\varphi_{00}\) is the only scalar mode that develops a lasting offset. We therefore compute the time-dependent memory-only detector response as
\begin{align}\label{eq:RCsGBMemoryResponse}
\Delta P_{\myst{RCsGB}}^{\rm mem}(t)
=&\,F_b(\Omega_d)\,\alpha\,Y_{00}(\Omega)\left[\varphi_{00}(t)-\varphi^I_{00}\right]\nonumber\\
&+F_+(\Omega_d)\,{}_{-2}Y_{20}(\Omega)\,\delta h^{\myst{sGB}}_{20}(t).
\end{align}
Here \(\varphi^I_{00}=\varphi_{00}(t_I)\) denotes the initial scalar monopole value, while \(\delta h^{\myst{sGB}}_{20}(t)\) is the tensor-memory mode shown in the lowest panel of Fig.~\ref{fig:Waveforms} that contributes to the $+$-polarization detector response.

The effect of scalar memory in the total memory response is illustrated in Fig.~\ref{fig:DetectorResponse}, where we compare the memory-only detector response for four representative source--detector configurations in GR, pure sGB, and RCsGB and also report the relative changes in total memory
\begin{equation}\label{eq:RelDevTotMemory} 
    \Delta_{\myst{sGB}}\equiv \frac{\Delta P_{\myst{sGB}}^{\myst{tot}}-\Delta P_{\myst{GR}}^{\myst{tot}}}{\Delta P_{\myst{GR}}^{\myst{tot}}}\,,\; \Delta_{\myst{RCsGB}}\equiv \frac{\Delta P_{\myst{RCsGB}}^{\myst{tot}}-\Delta P_{\myst{GR}}^{\myst{tot}}}{\Delta P_{\myst{GR}}^{\myst{tot}}}.
\end{equation}
Throughout, we take the benchmark value \(\alpha=\sqrt{10^{-5}}\) for the non-minimal Ricci coupling parameter, corresponding to the current upper limit from the strongest weak-field constraints discussed in Section~\ref{sSec:observationalConstraints}. The comparison between GR and pure sGB isolates the modification of the tensor memory induced by the GB dynamics alone, while the further comparison with RCsGB captures the additional effect of the scalar-memory contribution. Except for configurations with vanishing detector sensitivity to the scalar breathing mode, this additional contribution produces a further shift in the final memory offset of comparable observational size.

Importantly, the transition in scalar charge, and hence the buildup of the associated scalar memory, occurs on the same merger timescale as the nonlinear tensor memory, as can already be seen in Fig.~\ref{fig:Waveforms}. This is observationally relevant, because frequency-band-limited GW detectors are only sensitive to the memory \emph{transition} rather than to a constant offset. The scalar-memory contribution therefore enters the data in the same low-frequency band associated with the merger-driven rise of the nonlinear memory. Moreover, the sign of \(F_b\) in Eq.~\eqref{DPb} makes this scalar contribution constructive in the total memory response despite the decrease of the scalar monopole across merger.

\begin{figure}[h]
    \centering
    \includegraphics[width=1\linewidth]{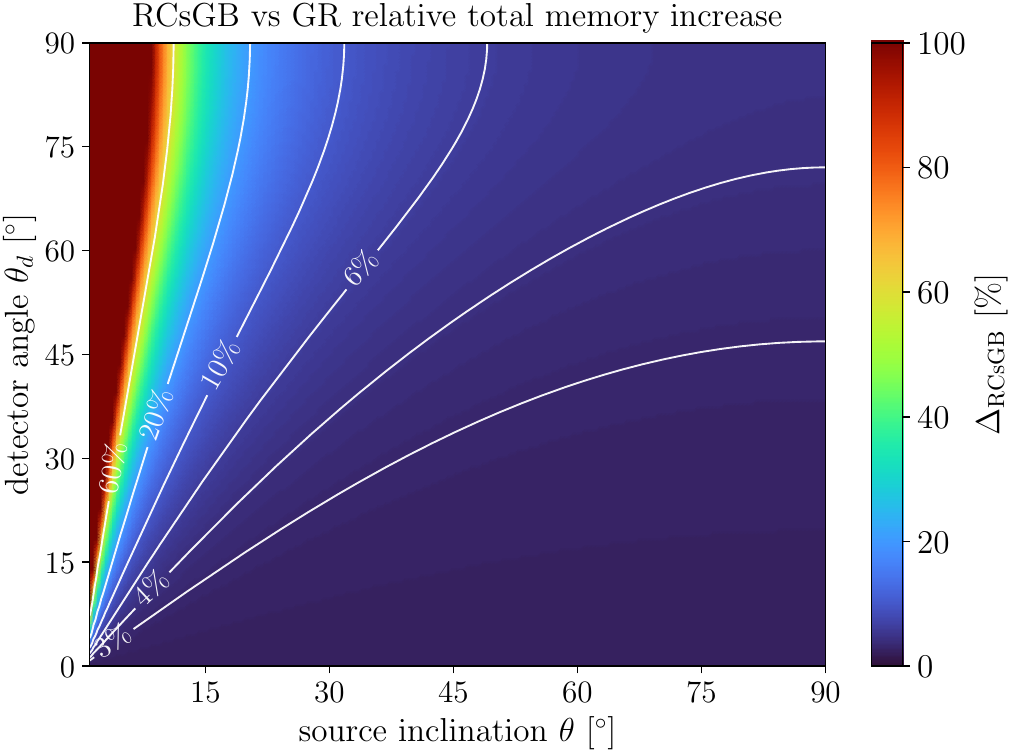}
    \caption{Final relative memory deviation in the detector response $\Delta_{\rm RCsGB}$ [Eq.~\eqref{eq:RelDevTotMemory}] as a function of the source inclination \(\theta\) and detector angle \(\theta_d\). For comparison, the corresponding pure sGB deviation of the memory amplitude $\Delta_{\rm sGB}$ stays approximately constant at \(2.2\%\) throughout this parameter space.}
    \label{fig:2DRelativeMemoryDeviation}
\end{figure}

More generally, the relative importance of the scalar memory increases towards face-on configurations because the tensor memory is inclination-suppressed, whereas the monopolar scalar contribution is nearly inclination independent. The face-on regime thus enhances the relative impact of scalar memory while reducing the absolute observability of the total memory signal, implying a nontrivial balance between fractional deviation and detectability. On the detector side, the effect is modulated by the breathing antenna pattern \(F_b\): the scalar memory impact remains substantial over a broad range of detector orientations and only disappears as \(\theta_d\to0\) for the single-detector response considered here. The resulting angular dependence is summarized in Fig.~\ref{fig:2DRelativeMemoryDeviation}.

We have also checked the dependence on mass ratio using waveform sets with \(q=1,2,3\) at fixed normalized coupling~\cite{AresteSalo:2025sxc}. As \(q\) increases, the emitted gravitational-wave energy decreases and the nonlinear tensor memory is reduced~\cite{Islam:2021old}, lowering the overall memory detectability~\cite{Inchauspe:2024ibs}. At the same time, the relative pure-sGB tensor-memory deviation decreases with \(q\), in agreement with Ref.~\cite{Gasparotto:2026bru}. In RCsGB, however, the scalar breathing contribution becomes a larger fraction of the beyond-GR correction to the detector response. We find that already for \(q=2\) the breathing channel accounts for more than two thirds of the beyond-GR correction for any detector orientation with \(\theta_d\gtrsim60^\circ\). Further details are given in Appendix~\ref{App:unequal}.

\section{A general beyond GR memory mechanism}

The RCsGB example discussed above points to a more general lesson. Observable memory from compact-binary coalescences is not controlled only by modifications of the tensor null-memory kernel. If a merger changes the asymptotic charge of an additional gravitational degree of freedom, and if that degree of freedom enters the physical detector response through an extra polarization channel, the charge reconfiguration can generate an additional memory contribution that competes directly with tensor-memory modifications.

\subsection{RCsGB charge relaxation}\label{sec:underRCsGB}

The RCsGB result can be understood directly from the structure of shift-symmetric sGB black-hole hair~\cite{Witek:2018dmd,AresteSalo:2025sxc}. In this class of theories, a black hole of ADM mass \(M\) carries a monopolar scalar charge whose asymptotic field scales as
\begin{equation}\label{eq:ScalarChargeScalingsGB}
    \varphi_{00}\simeq\frac{\lambda_\alpha}{rM}\, .
\end{equation}
For an initial binary with \(q=m_1/m_2\) and final remnant mass \(M\simeq m_1+m_2\), this implies
\begin{equation}
    \varphi^I_{00}\sim \frac{\lambda_\alpha}{r}
    \left(\frac{1}{m_1}+\frac{1}{m_2}\right),
    \qquad
    \varphi^F_{00}\sim \frac{\lambda_\alpha}{rM}.
\end{equation}
The ratio between the initial and final monopolar scalar charges is therefore
\begin{equation}\label{eq:scalarratio}
    \frac{\varphi^I_{00}}{\varphi^F_{00}}
    \simeq \frac{(1+q)^2}{q}
    =\frac{1}{\eta}\, .
\end{equation}
This explains the scalar-charge drop observed in Fig.~\ref{fig:Waveforms}. The scalar breathing memory is not controlled by scalar radiative flux, but by the merger-driven transition from two scalar-charged black holes to a single remnant in a theory where the scalar charge is not conserved. Since this reconfiguration occurs on the merger timescale, the scalar breathing memory enters the detector response through the same low-frequency transition as tensor memory. The corresponding mass-ratio dependence is discussed in Appendix~\ref{App:unequal}.

\subsection{Conjecture of generality}\label{sSec:GeneralityConjecture}

The mechanism suggested by RCsGB should extend beyond this specific theory. The essential ingredients are:
\begin{enumerate}[(i)]
    \item the theory contains an additional gravitational degree of freedom with a radiative component;
    \item compact objects carry a state-dependent charge under that field, fixed by the equilibrium solution rather than protected by an exact Noether conservation law;\footnote{If the extra charge were exactly conserved, any monopole jump would have to be carried by flux to infinity and would typically be much smaller. The large effect arises when the merger instead relaxes to a remnant whose equilibrium charge differs substantially from the sum of the progenitor charges.}
    \item the corresponding perturbation enters the physical detector response through an observable polarization channel.
\end{enumerate}
Under these conditions, the detector memory acquires a schematic contribution
\begin{equation}\label{eq:Conjecture}
    \Delta P_{\Lambda}\sim F_{\Lambda}(\Omega_d)\,\alpha_{\Lambda}\,\Delta Q_{\Lambda}\,,
\end{equation}
where \(Q_{\Lambda}\) is the asymptotic charge of the extra field, \(\alpha_{\Lambda}\) is its effective coupling to the detector response, and \(F_{\Lambda}\) is the corresponding antenna pattern. Such a memory contribution is manifestly distinct from tensor null memory, which is tied to radiative energy flux.

This reasoning applies naturally to theories with hairy compact objects and observable extra polarizations, including broad sectors of scalar-tensor gravity.\footnote{For a scalar linearly sourced by a curvature invariant with \(2n\) derivatives, one expects \(\varphi_{00}\sim \lambda_\alpha M^{3-2n}/r\), reproducing the sGB scaling in Eq.~\eqref{eq:ScalarChargeScalingsGB} for \(n=2\). The same mechanism can also operate when the charge does not follow such a power law, as in nonperturbative scalarization models, provided the merger changes the equilibrium scalar charge~\cite{Doneva:2022ewd,East:2021bqk,Silva:2020omi,Elley:2022ept,Julie:2023ncq,Capuano:2026lhs,Doneva:2021tvn}.} 
The mechanism can also arise in theories that satisfy a black-hole no-hair theorem. A typical example is provided by massless Brans-Dicke-type scalar-tensor theories, where stationary black holes carry no scalar charge, while neutron stars may carry a nonzero scalar monopole through spontaneous scalarization. If such a compact binary merger ends in a black-hole remnant, the initial scalar monopole of the system can be radiated away or collapse to zero, generating scalar memory in the breathing channel. This is the compact-binary analogue of the scalar memory generated by no-hair transitions in gravitational collapse discussed in Ref.~\cite{Du:2016hww}. A particularly suggestive example is provided by black-hole-neutron-star mergers of scalarized neutron stars, as considered in Ref.~\cite{Ma:2023sok}.

\section{Conclusion}\label{Sec:Conclusion}

Any theory satisfying the conditions (i)--(iii) identified in Section~\ref{sSec:GeneralityConjecture} should exhibit the same qualitative phenomenon: a merger-induced change in the charge of an additional gravitational degree of freedom, when combined with an observable extra polarization channel, can produce a potentially leading beyond-GR correction to the memory signal.

We have demonstrated this mechanism explicitly in Ricci-coupled scalar-Gauss-Bonnet gravity. In contrast to standard sGB gravity, the Ricci coupling makes the scalar degree of freedom directly observable as a breathing polarization in the physical detector response. Using numerical-relativity waveforms from the corresponding vacuum shift-symmetric sGB theory, we showed that the change in scalar charge across merger generates an additional scalar-memory contribution in the breathing channel. For a GW150914-like system and values of the Ricci coupling compatible with current weak-field bounds, this contribution modifies the observable memory by an amount comparable to the pure-sGB correction to tensor memory, and can substantially enhance the total beyond-GR deviation for favorable source and detector configurations.

This has two immediate implications. First, beyond-GR memory from compact-binary coalescences cannot, in general, be characterized only by modifications of tensor null memory. Once additional polarizations are present, the corresponding scalar memory must also be included in the detector response. Second, because scalar breathing memory has a distinct dependence on binary inclination and detector orientation, it provides a complementary low-frequency probe of additional gravitational degrees of freedom.

The scalar-memory contribution emphasized here is conceptually distinct from tensor null memory: it is tied to a change in a non-conserved scalar charge rather than to a radiative energy-flux reconstruction. Clarifying its relation to asymptotic symmetries and soft theorems, and incorporating it into detector forecasts and waveform models, are important directions for future work.

\begin{acknowledgments}
JZ is supported by funding from the Swiss National Science Foundation (Grant No. 222346) and the Janggen-Pöhn-Foundation. The Center of Gravity is a Center of Excellence funded by the Danish National Research Foundation under Grant No. 184. LAS is partly funded by Interuniversitaire Bijzonder Onderzoeksfonds (IBOF)/21/084. DD acknowledges financial support from
the Spanish Ministry of Science and Innovation through the Ram\'on y Cajal programme (grant RYC2023-042559-I), funded by MCIN/AEI/10.13039/501100011033 and by ESF+, from an Emmy Noether Research Group funded by the German Research Foundation (DFG) under Grant No. DO 1771/1-1,
and by the Spanish Agencia Estatal de Investigacion (grant PID2024-159689NB-C21) funded by MICIU/AEI/10.13039/501100011033 and by FEDER / EU. SY is supported by the European
Union-NextGenerationEU, through the National Recovery and Resilience Plan of the Republic of Bulgaria,
project No. BG-RRP-2.004-0008-C01. We acknowledge Discoverer PetaSC and EuroHPC JU for awarding this project access to Discoverer supercomputer resources.
\end{acknowledgments}

\appendix
\section{Embedding into Horndeski gravity}\label{App:Horndeski}

The action of Horndeski theory as the most general scalar-tensor theory with second order equations of motion can be written as
\cite{Horndeski:1974wa,Nicolis:2008in,Deffayet:2009wt,Deffayet:2009mn,Heisenberg:2018vsk,Kobayashi:2019hrl}
\begin{equation}\label{eq:ActionHorndeski}
    S^{\myst{H}}=\frac{1}{16\pi}\int d^4 x\sqrt{-g}\left(\sum_{i=2}^5L^{\myst{H}}_i\right)+S_\text{m}[g,\Psi_\text{m}]\,,
\end{equation}
where
\begin{align}
        L^{\myst{H}}_2=&\,G_2(\Phi,X)\,,\\
        L^{\myst{H}}_3=&-G_3(\Phi,X)\Box\Phi\,,\\
        L^{\myst{H}}_4=&\,G_4(\Phi,X)\,R+G_{4,X}\left[(\Box\Phi)^2-\Phi^{\mu\nu}\Phi_{\mu\nu}\right]\,,\\
        L^{\myst{H}}_5=&\,G_5(\Phi,X)\,G^{\mu\nu}\Phi_{\mu\nu}-\frac{G_{5,X}}{6}\Big[(\Box\Phi)^3\nonumber\\
        &-3\,\Box\Phi\,\Phi^{\mu\nu}\Phi_{\mu\nu}+2\,\Phi_{\mu\nu}\Phi^{\nu\lambda}\Phi\du{\lambda}{\mu}\Big]\,,
\end{align}
with $\Phi_{\mu\nu}\equiv\nabla_{\mu}\nabla_{\nu}\Phi$, and where the $G_i$'s are arbitrary functionals of $\Phi$ and the kinetic combination 
\[
    X\equiv - \frac{1}{2} \partial_\mu\Phi\partial^\mu\Phi\,.
\]
Moreover, we use the standard notation $G_{i,Z}\equiv\partial G_i/\partial Z$.
This theory in particular also encompasses Ricci-coupled sGB gravity considered in this work for specific choices of the functionals $G_i$. Concretely, the action in Eq.~\eqref{eq:ActionsString} is recovered through
\begin{subequations}\label{eq:CorrespondencesGBHorndeski}
\begin{align}
        G_2&=-f(\Phi)\omega(\Phi) X-8\lambda f^{(4)}(\Phi)X^2(3-\ln X)\,,\\ G_3&=-4\lambda f^{(3)}(\Phi)X(7-3\ln X)\,,\\
        G_4&=f(\Phi)-4\lambda f^{(2)}(\Phi)X(2-\ln X)\,,\\
        G_5&=4\lambda f^{(1)}(\Phi)\ln X\,,
\end{align}
\end{subequations}
where $f^{(n)}(\Phi)\equiv\partial^n f/\partial\Phi^n$. 

\subsection{Full results in Horndeski gravity}

For Horndeski theory, the gravitational polarizations~\cite{Hou:2017bqj,Heisenberg:2024cjk} and the corresponding memory formulas~\cite{Heisenberg:2024cjk,Heisenberg:2023prj,Zosso:2024xgy,Gasparotto:2026bru} in asymptotically flat spacetime have been computed in full generality. The Ricci-coupled sGB results used in the main text are recovered from these expressions by applying the correspondences in Eq.~\eqref{eq:CorrespondencesGBHorndeski} within the asymptotic limit. For completeness, we briefly recall these results in this section.

\subsubsection{Polarizations}
In Horndeski gravity the response matrix in the polarization space [Eq.~\eqref{eq:Pijexpand}] is given by
\cite{Hou:2017bqj,Heisenberg:2024cjk,Zosso:2024xgy}
\begin{align}\label{eq:PijexpandApp}
P_{ij}=e^+_{ij}\,h_++e^\times_{ij}\,h_\times-\sigma \,e^b_{ij}\,\varphi_1+\sigma\left(v_\text{S}^2-1\right)e^l_{ij}\varphi_1\,,
\end{align} 
where
\begin{equation}\label{eq:DefSigma}
    \sigma \equiv \frac{\bar{G}_{4,\Phi}}{\bar{G}_4}\,,\quad v_\text{S}(\omega)=\sqrt{1-\frac{m^2}{\omega^2}}\,,
\end{equation}
and
\begin{align}\label{eq:DefMass SVH}
    m^2 = \frac{\bar{G}_{2,\Phi\Phi}}{\bar{G}_{2,X}-2\bar{G}_{3,\Phi}+3\sigma^2\bar{G_4} }\,. 
\end{align}
We employ here the notation $\bar{G}_i\equiv G_i(\varphi_0,0)$ to denote an evaluation on the asymptotically flat background. Hence, the three degrees of freedom of the theory excite at most four gravitational polarizations, where the polarization tensor of the additional longitudinal scalar polarization simply reads
\begin{align}\label{eq:PolTensorsLongitudinlal}
e^l_{ij} &\equiv n_in_j\,.
\end{align}

\subsubsection{Memory}

In massless Horndeski theory, the tensor null-memory contribution sourced by gravitational radiation takes the form of Eq.~\eqref{eq:NonLinDispMemorysGB}, but with the scalar energy flux given by
\begin{equation}\label{eq:EperSolidAngleHorndeski}
    F_{\myst{H}}(u,\Omega')
    \equiv
    \int_{-\infty}^u d u' \, r'^2
    \left\langle |\dot h|^2+\rho^2\dot{\varphi}_1^{\,2}\right\rangle ,
\end{equation}
where
\begin{equation}\label{eq:DefRho}
    \rho^2\equiv
    \frac{\bar G_{2,X}-2\,\bar G_{3,\Phi}}{\bar G_4}
    +3\,\frac{\bar G_{4,\Phi}^2}{\bar G_4^2}\, .
\end{equation}
Thus, \(\rho\) controls the scalar-radiation contribution to tensor null memory.

\subsection{Recovering Ricci-coupled sGB results}

RCsGB gravity is recovered through Eqs.~\eqref{eq:CorrespondencesGBHorndeski}.
Evaluated on the asymptotically flat background we obtain [Eq.~\eqref{eq:DefMass SVH}]
\begin{equation}
    m=0\; \Leftrightarrow \; v_\text{S}=1\,.
\end{equation}
This implies that, as expected, the scalar DOF in RCsGB is massless due to the absence of a scalar potential and the longitudinal polarization in Eq.~\eqref{eq:PijexpandApp} is absent. The transverse breathing response is controlled by [Eq.~\eqref{eq:DefSigma}]
\begin{equation}
    \sigma
    =
    \frac{f'(\varphi_0)}{f(\varphi_0)}\,,
\end{equation}
while the scalar contribution to tensor null memory is characterized by [Eq.~\eqref{eq:DefRho}]
\begin{equation}
    \rho^2
    =
    \omega(\varphi_0)+3\sigma^2\,.
\end{equation}
For the exponential coupling in Eq.~\eqref{eq:nonminimalcoupling1},
\begin{equation}
    \sigma=-\alpha\,,
    \qquad
    \rho^2=\omega(\varphi_0)+3\alpha^2\,.
\end{equation}
With the Einstein-frame canonical normalization in Eq.~\eqref{eq:EinsteinFrameCanonical_omega}, this reduces to
\begin{equation}
    \rho^2=1\,.
\end{equation}
Consequently, Eqs.~\eqref{eq:PijexpandApp} and \eqref{eq:EperSolidAngleHorndeski} reduce to the main-text results in Eq.~\eqref{eq:Pijexpand} and \eqref{eq:EperSolidAngleGR} respectively.

In contrast, standard shift-symmetric sGB gravity has no non-minimal Ricci coupling in the physical metric sector, corresponding to \(\sigma=0\). The scalar field can then affect the vacuum dynamics and source tensor null memory through its radiative energy flux, but it does not enter the detector response as an independent breathing polarization.

\section{Weyl transformation to Einstein frame}\label{App:Weyl details}

In this section, we summarize the transformation from the physical Jordan frame to the Einstein frame for a general non-minimal coupling function \(f(\Phi)\). We begin from the Jordan-frame action in Eq.~\eqref{eq:ActionsString}, where \(g_{\mu\nu}\) is the physical metric minimally coupled to matter. We assume \(f(\Phi)>0\), so that the Weyl rescaling
\begin{equation}
    \tilde g_{\mu\nu}=f(\Phi)\,g_{\mu\nu},
    \qquad
    g_{\mu\nu}=f(\Phi)^{-1}\tilde g_{\mu\nu}
\end{equation}
is non-singular in the field range of interest.\footnote{More generally, we assume that \(f\) and \(\omega\) are regular in the field range of interest, with \(f>0\) so that the effective Planck mass has the correct sign. We furthermore require the Einstein-frame scalar kinetic coefficient \(\Omega(\Phi)\) in Eq.~\eqref{eq:EinsteinFrameKineticTerm} to remain positive, so that no ghostlike scalar mode propagates.}

In four dimensions,
\begin{equation}
    \sqrt{-g}=f^{-2}\sqrt{-\tilde g},
    \qquad
    g^{\mu\nu}=f\,\tilde g^{\mu\nu},
\end{equation}
and the Ricci scalar transforms as
\begin{equation}
    R
    =
    f(\Phi)\left[
        \tilde R
        -6\tilde \Box \Sigma
        -6(\tilde\nabla\Sigma)^2
    \right],
    \qquad
    \Sigma\equiv-\frac12\ln f .
\end{equation}
Discarding the total derivative \(\tilde\Box\Sigma\), the Ricci and scalar-kinetic sectors become
\begin{equation}
    \sqrt{-g}\,f(\Phi)
    \left[
        R+\omega(\Phi)X
    \right]
    \hat{=}
    \sqrt{-\tilde g}
    \left[
        \tilde R+\Omega(\Phi)X
    \right],
\end{equation}
with
\begin{align}\label{eq:EinsteinFrameKineticTerm}
    \Omega(\Phi)\equiv \omega(\Phi)+3\left(\frac{f'(\Phi)}{f(\Phi)}\right)^2 .
\end{align}
Here and in the following \(X\) denotes the scalar kinetic term formed with the metric used in the corresponding frame.

The Gauss--Bonnet term transforms as
\begin{equation}
    \mathcal G[g]
    =
    f(\Phi)^2
    \left[
        \tilde{\mathcal G}
        +
        \Delta_{\rm conf}
    \right],
\end{equation}
where \(\Delta_{\rm conf}\) contains terms with derivatives of \(\Sigma=-\tfrac12\ln f\), and hence derivatives of the scalar field. In the perturbative weak-coupling regime relevant here, these Weyl-generated derivative interactions are subleading at the order considered in the main text. Indeed, for solutions continuously connected to GR, the scalar profile is coupling-induced and schematically satisfies \(\Phi\sim\lambda_\alpha/M^2\). The terms in \(\Delta_{\rm conf}\) are therefore suppressed by additional powers of the weak-coupling expansion parameter relative to the leading \(f(\Phi)\tilde{\mathcal G}\) contribution. We consequently keep only this leading term, assuming \(\lambda_\alpha/M^2\ll1\). For the small-field expansion \(f_\alpha(\Phi)=e^{-\alpha\Phi}\simeq1-\alpha\Phi\), we also assume \(|\alpha\Phi|\ll1\).

To leading order in this expansion, the Einstein-frame action takes the form
\begin{align}\label{eq:EinsteinFrameGeneralApp}
    S_E
    \simeq&\, \frac{1}{16\pi}
    \int d^4x \sqrt{-\tilde g}
    \left[
        \tilde R
        +
        \Omega(\Phi)X
        -
        \lambda f(\Phi)\tilde{\mathcal G}
    \right]
    \nonumber\\
    &+S_\text{m}[f(\Phi)^{-1}\tilde g_{\mu\nu},\Psi_\text{m}]\, .
\end{align}
Provided \(\Omega(\Phi)>0\), one may canonically normalize the Einstein-frame scalar by a field redefinition.
In the main text, for simplicity, we work in a parametrization in which this canonical normalization is already realized with \(\Omega(\Phi)=1\). This choice is justified because it does not alter the leading-order binary-black-hole dynamics, but merely fixes the normalization convention for the scalar field and hence for the effective Gauss--Bonnet and matter couplings. Finally, since the scalar field approaches a constant value \(\varphi_0\) at spatial infinity, which can be normalized such that \(f(\varphi_0)=1\), the Einstein- and Jordan-frame ADM masses coincide in the convention used throughout this work.

For the exponential coupling in Eq.~\eqref{eq:nonminimalcoupling1}, we therefore obtain 
\begin{equation}\label{eq:EinsteinFrameCanonical_omega}
    \omega_\alpha(\Phi)=1-3\alpha^2\, .
\end{equation}
and Eq.~\eqref{eq:EinsteinFrameGeneralApp} reduces to the leading-order Einstein-frame action in Eq.~\eqref{eq:actionEinstein}.

\section{Geometric conventions and detector response}
\label{App:Definitions}
\subsection{Radiation basis and polarization tensors}\label{sApp:RadiationBasis}

Given a source-centered spherical coordinate system $\{t,r,\Omega=(\theta,\phi)\}$ the unit radial vector reads
\begin{equation}\label{eq:Def Direction of Propagation n}
    \mathbf{n}(\Omega)=(\sin\theta \cos\phi,\,\sin\theta \sin\phi,\,\cos\theta )\,.
\end{equation}
The associated transverse space of a given direction $\mathbf{n}(\Omega)$ can then be described through 
\begin{widetext}
\begin{subequations}\label{eq:vectorbasispGen}
\begin{align}
&\bm{\theta}(\Omega,\psi)=(\cos\theta \cos\phi \cos\psi+\sin\phi\sin\psi,\,\cos\theta \sin\phi \cos\psi-\cos\phi \sin\psi,\,-\sin\theta\cos\psi)\,,\\
&\bm{\phi}(\Omega,\psi)=(\cos\theta \cos\phi \sin\psi-\sin\phi\cos\psi,\,\cos\theta\sin\phi\sin\psi+ \cos\phi \cos\psi,\,-\sin\theta\sin\psi)\,,
\end{align}
\end{subequations}
\end{widetext}
where $\psi$ represents the polarization angle that captures the freedom of rotating the transverse basis about the direction of propagation $ \mathbf{n}$.
These spatial vectors form an orthonormal basis satisfying the completeness relation
\begin{equation}\label{eq:polcompleteness}
\delta_{ij}=n_in_j+\theta_i\theta_j+\phi_i\phi_j\,.
\end{equation} 
For the choice \(\psi=0\), the transverse vectors \(\bm{\theta}\) and \(\bm{\phi}\) coincide with the natural polar and azimuthal basis directions,
\begin{subequations}\label{eq:Def Transverse Vectors u v}
\begin{align}
\bm{\theta}&=(\cos\theta \cos\phi,\,\cos\theta \sin\phi,\,-\sin\theta )\,,\\
\bm{\phi}&=(-\sin\phi,\,\cos\phi,\,0)\,.
\end{align}
\end{subequations}
For the non-precessing binaries considered here, with angular momentum aligned with the \(z\)-axis, this polarization basis places the full amplitude of the linearly polarized memory signal in the \(+\)-polarization, with no cross-polarized memory.

\begin{figure}[h]
    \centering
    \subfloat[$|F_{+}|$\label{fig:antenna_plus}]{
        \includegraphics[width=0.145\textwidth]{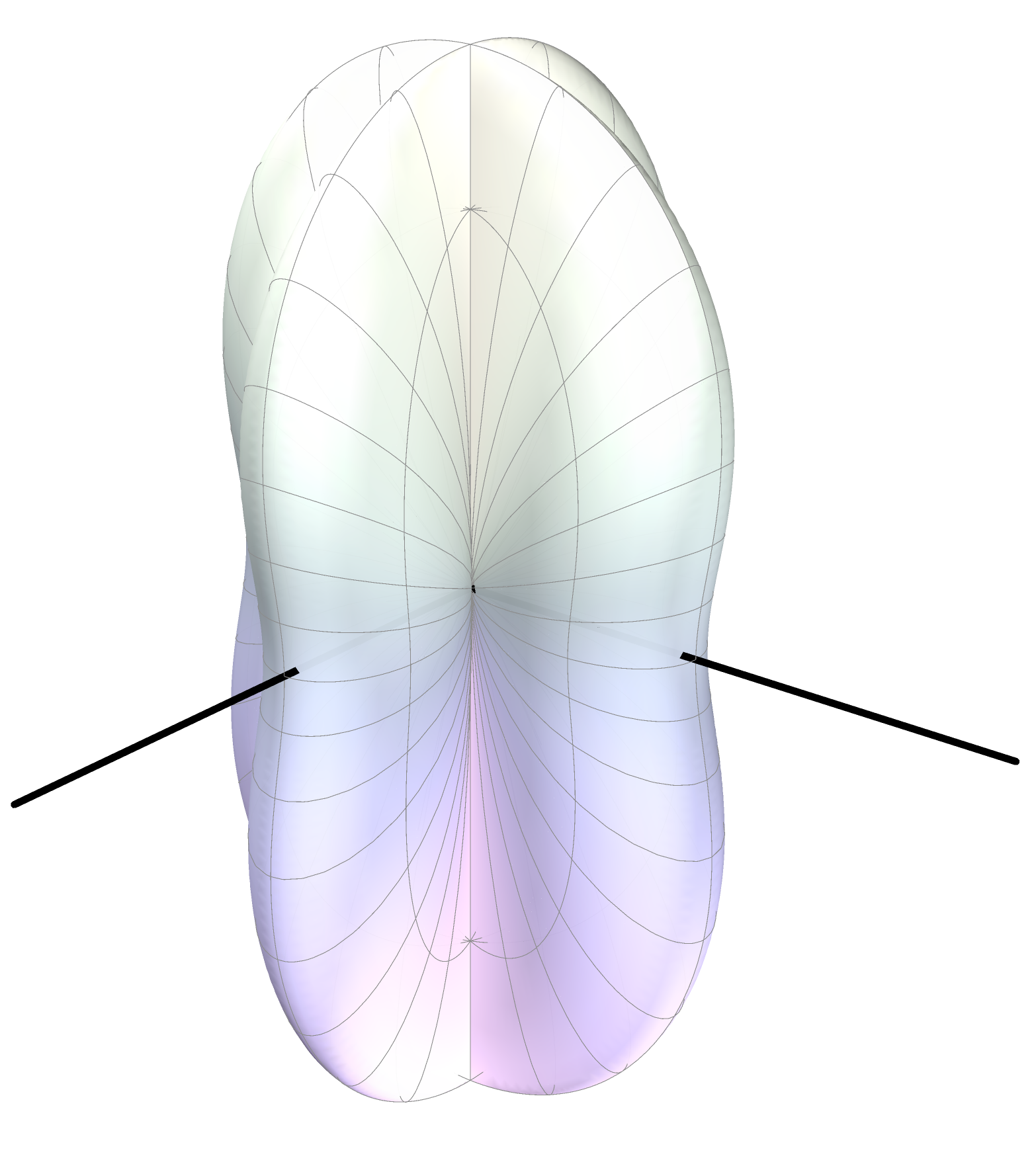}
    }
    \hfill
    \subfloat[$|F_{\times}|$\label{fig:antenna_cross}]{
        \includegraphics[width=0.145\textwidth]{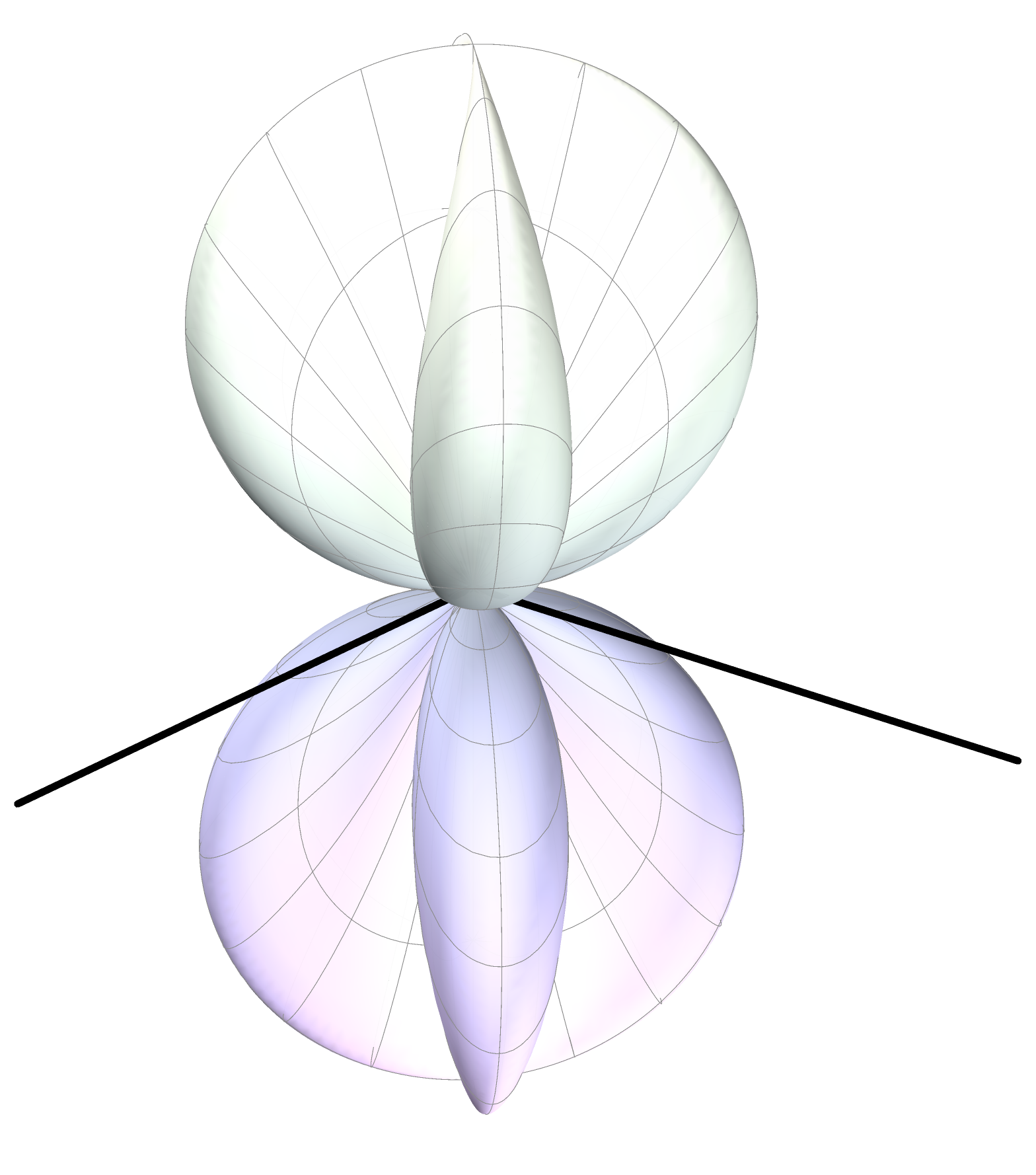}
    }
    \hfill
    \subfloat[$|F_{b}|$\label{fig:antenna_breathing}]{
        \includegraphics[width=0.145\textwidth]{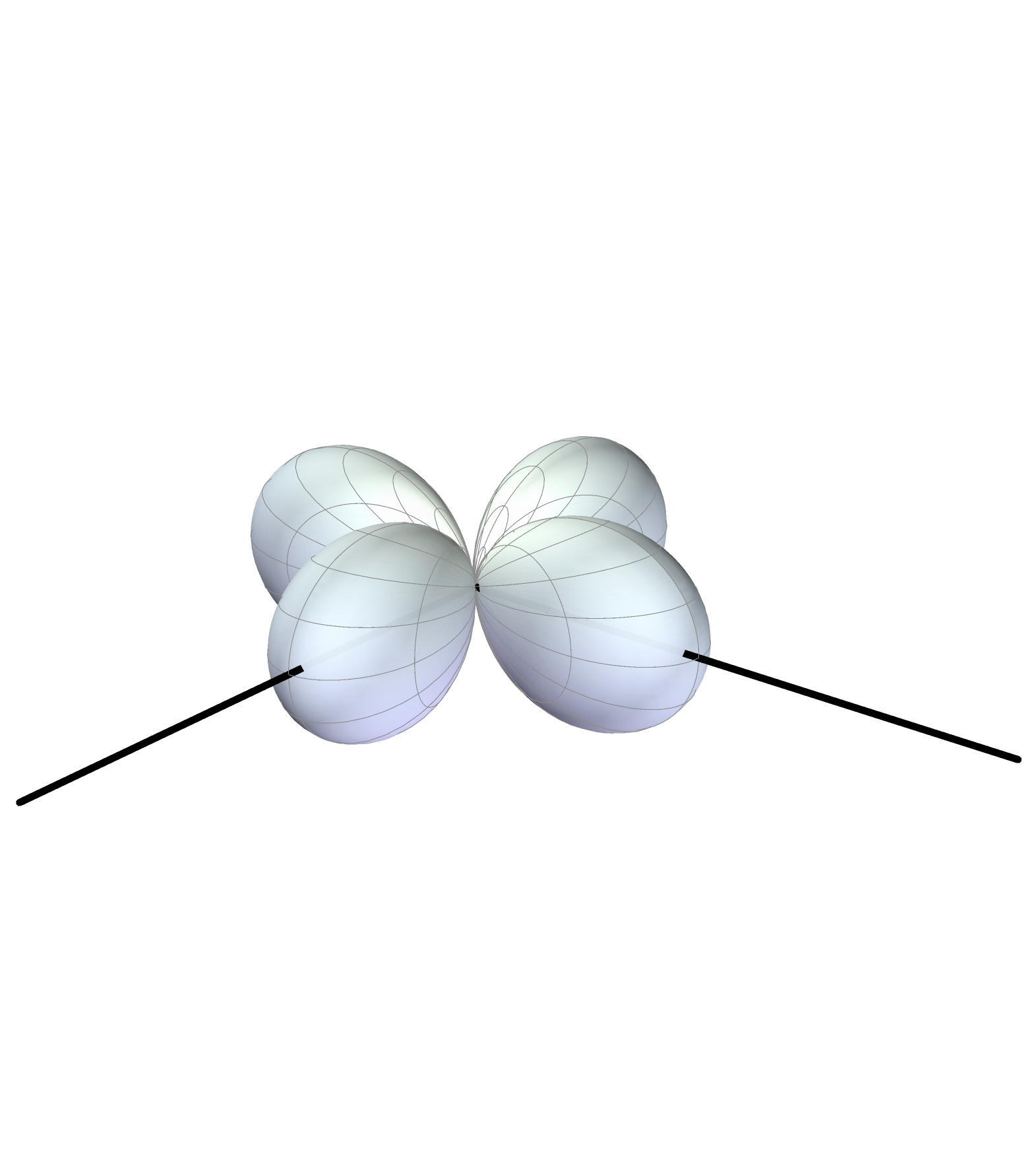}
    }
    \caption{Absolute values of the antenna pattern functions \(|F_\lambda|\) for the two tensor polarizations and the scalar breathing polarization, shown as functions of the source direction \((\theta_d,\phi_d)\) for \(\psi=0\). The radial distance represents the angular response to a unit-amplitude GW of an ideal interferometer with detector arms along the \(x_d\) and \(y_d\) directions, indicated by the black lines.}
    \label{fig:antenna_patterns}
\end{figure}

In terms of this orthonormal basis of the asymptotic radiation \(\{n_i,\theta_i,\phi_i\}\), the three independent polarization tensors in Eq.~\eqref{eq:Pijexpand} can be written as
\begin{subequations}\label{eq:PolTensors}
\begin{align}
e^+_{ij}&=\theta_i\theta_j-\phi_i\phi_j\,,
\qquad
e^\times_{ij}=\theta_i\phi_j+\phi_i\theta_j\,,
\\
e^b_{ij}&\equiv \theta_i\theta_j+\phi_i\phi_j
=\delta_{ij}-n_in_j\,.
\end{align}
\end{subequations}

\subsection{Detector pattern functions}\label{sApp:AntennaPattern}
For an equal-arm interferometer with arms along the detector-frame directions
\(\mathbf{e}^{x_d}\) and \(\mathbf{e}^{y_d}\), the detector response to the
polarization matrix \(P_{ij}\) in Eq.~\eqref{eq:integratedGeodesicDeviation} is
\begin{equation}
P(t)=\frac{1}{2}\left(e^{x_d}_ie^{x_d}_j-e^{y_d}_ie^{y_d}_j\right)P^{ij}\,.
\end{equation}
The response to each individual polarization is encoded in the corresponding
\emph{antenna pattern function}
\begin{equation}\label{eq:defDetectorPattern}
F_\Lambda\equiv \frac{1}{2}\left(e^{x_d}_ie^{x_d}_j-e^{y_d}_ie^{y_d}_j\right)e_\Lambda^{ij}\,.
\end{equation}
These functions describe the directional sensitivity of the detector to the
different GW polarizations in Eq.~\eqref{eq:PolarizationMatrixDef}.

Writing the direction of GW propagation in detector-frame spherical angles
\(\Omega_d=\{\theta_d,\phi_d\}\), we choose
\begin{equation}\label{eq:DefDirectionPropagationDetector}
    \mathbf{n}(\Omega_d)=-(\sin\theta_d \cos\phi_d,\,
    \sin\theta_d \sin\phi_d,\,
    \cos\theta_d)\,,
\end{equation}
where the minus sign reflects that the wave propagates \emph{towards} the
detector from the source direction \((\theta_d,\phi_d)\). Using the
polarization tensors in Eq.~\eqref{eq:PolTensors}, constructed from the
radiation basis vectors of Eq.~\eqref{eq:vectorbasispGen}, one obtains the
standard antenna pattern functions for an L-shaped interferometer
\begin{subequations}\label{eq:DetectorPatternApp}
\begin{align}
F_+&=\frac{1}{2}\left(1+\cos^2\theta\right)\cos2\phi\cos2\psi+\cos\theta\sin2\phi\sin2\psi,\\
F_\times&=\frac{1}{2}\left(1+\cos^2\theta\right)\cos2\phi\sin2\psi-\cos\theta\sin2\phi\cos2\psi,\\
F_{b}&=-\frac{1}{2}\sin^2\theta\cos2\phi\,.\label{DPb}
\end{align}
\end{subequations}
Note that for the breathing mode the dependence on the polarization angle \(\psi\) drops out, as expected for a scalar polarization. 

This detector response assumes the standard long-wavelength and rigid-detector limit, in which the arm length is much smaller than the GW wavelength and the detector can be treated as an instantaneous differential strain meter. We also neglect the detector's rotation during the signal, which is an excellent approximation for burst-like BBH merger signals and for the memory buildup considered here. The absolute values of the antenna patterns are shown in Fig.~\ref{fig:antenna_patterns} for \(\psi=0\), the convention adopted in the main text.

\section{Mass-ratio dependence}\label{App:unequal}

We collect here the mass-ratio dependence of the memory response for the unequal-mass simulations with \(q=1,2,3\) at fixed \(\lambda_\alpha/m_2^2=0.106\) \cite{AresteSalo:2025sxc}. To quantify the importance of the breathing channel within the beyond-GR correction, we define the absolute fractional scalar-breathing weight
\begin{equation}\label{eq:RelImpScalarMem}
    \delta_b\equiv 
    \frac{|\Delta_{\rm RCsGB}-\Delta_{\rm sGB}|}
    {|\Delta_{\rm sGB}|+|\Delta_{\rm RCsGB}-\Delta_{\rm sGB}|}\, .
\end{equation}
The relative deviations \(\Delta_{\rm sGB}\) and \(\Delta_{\rm RCsGB}\) are defined in Eq.~\eqref{eq:RelDevTotMemory}.

The results are shown in Table~\ref{tab:MassRatioMemory} for the two representative geometries already employed in Fig.~\ref{fig:DetectorResponse}. Increasing \(q\) reduces the overall memory scale and suppresses the relative pure-sGB tensor-memory deviation \(\Delta_{\rm sGB}\), in agreement with Ref.~\cite{Gasparotto:2026bru}. A source decomposition of the tensor memory shows that, near equal mass, the scalar-radiation-sourced tensor-memory term, which contributes to the memory amplitude with negative sign, is negligible compared to the tensor-radiation contribution, as shown in Fig.~\ref{fig:Waveforms}. 
By \(q=3\), however, the scalar-radiation-induced tensor memory becomes comparable to the residual tensor-sector correction.
The slightly negative pure-sGB entry at \(q=3\) should, however, not be overinterpreted: it is a sub-percent value, close to the expected numerical-error level of the underlying sGB waveforms~\cite{Gasparotto:2026bru}. The main point of Table~\ref{tab:MassRatioMemory} is that, in RCsGB, the scalar breathing contribution becomes a larger fraction of the beyond-GR detector-response correction as \(q\) increases, as indicated by the larger values of the scalar breathing mode weight \(\delta_b\) for the two representative geometries with optimal breathing sensitivity, \(\theta_d=90^\circ\). 

\begin{table}[t]
\centering
\setlength{\tabcolsep}{5.2pt}
\begin{tabular}{@{}c ccc ccc@{}}
\toprule
\(q\) 
& \multicolumn{3}{c}{\(\theta=90^\circ,\ \theta_d=90^\circ\)}
& \multicolumn{3}{c}{\(\theta=10^\circ,\ \theta_d=90^\circ\)} \\
\cmidrule(lr){2-4}\cmidrule(l){5-7}
& \(\Delta_{\rm sGB}\) 
& \(\Delta_{\rm RCsGB}\) 
& \(\delta_b\)
& \(\Delta_{\rm sGB}\) 
& \(\Delta_{\rm RCsGB}\) 
& \(\delta_b\) \\
\midrule
1 & 1.9   & 3.7  & 50 & 1.9   & 64 & 97 \\
2 & 0.18  & 1.5  & 88 & 0.18   & 43 & 99 \\
3 & -0.38 & 0.94 & 78 & -0.38 & 43 & 99 \\
\bottomrule
\end{tabular}
\caption{
Mass-ratio dependence of the final memory-only detector response for two representative geometries. All entries are percentages. The quantities \(\Delta_{\rm sGB}\) and \(\Delta_{\rm RCsGB}\) are defined in Eq.~\eqref{eq:RelDevTotMemory}, while \(\delta_b\) is defined in Eq.~\eqref{eq:RelImpScalarMem}.
}
\label{tab:MassRatioMemory}
\end{table}

\begin{figure}[h]
    \centering
    \includegraphics[width=0.99\linewidth]{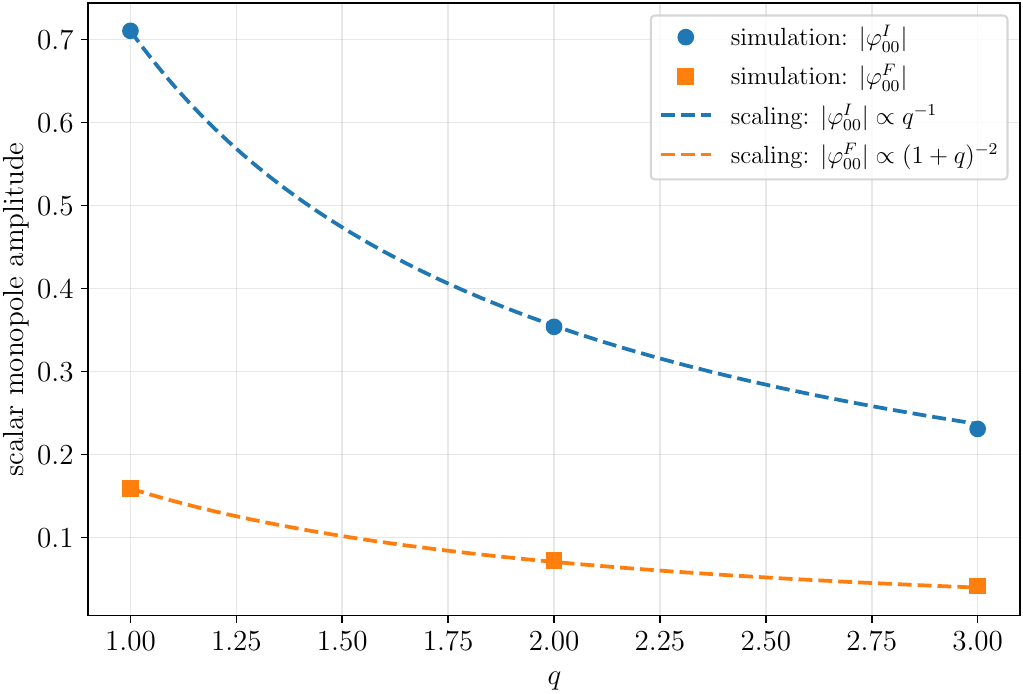}
    \caption{
    Initial and final scalar monopole amplitudes as functions of mass ratio at fixed total mass and fixed \(\lambda_\alpha/m_2^2\). Markers denote numerical values extracted from the scalar waveform. Smooth curves show the analytic scalings in Eq.~\eqref{eq:AppScalings} normalized to the corresponding \(q=1\) value.
    }
    \label{fig:scalarcharge_scaling}
\end{figure}

Fig.~\ref{fig:scalarcharge_scaling} displays the corresponding scalar monopole amplitudes. The markers are extracted from the \(\varphi_{00}\) waveform before and after merger. The smooth curves show the analytic expectations for the initial (inspiral) and final (post-merger) monopole amplitudes, discussed around Eq.~\eqref{eq:scalarratio} and normalized to the corresponding \(q=1\) numerical value:
\begin{equation}\label{eq:AppScalings}
    |\varphi_{00}^{I}|\propto q^{-1},
    \qquad
    |\varphi_{00}^{F}|\propto (1+q)^{-2}.
\end{equation}
The agreement supports the interpretation in Section~\ref{sec:underRCsGB} as a qualitative understanding of the results in Table~\ref{tab:MassRatioMemory}. Increasing \(q\) decreases the absolute scalar monopole amplitudes, but enhances the ratio between the initial and final scalar charge (see Eq.~\eqref{eq:scalarratio}). This explains why the scalar breathing memory can become relatively more important than the residual pure-sGB tensor-memory deviation at larger mass ratio.\footnote{Interestingly, in the large-\(q\) limit the absolute scalar jump scales as \(|\Delta\varphi_{00}|\sim q^{-1}\), and therefore even decreases more slowly with \(q\) than the total tensor memory scaling \(\Delta h\sim\eta^{1.65}\sim q^{-1.65}\) found in Ref.~\cite{Liu:2021zys}.}

\newpage
\bibliographystyle{utcaps}
\bibliography{refs}

@article{Liu:2021zys,
    author = "Liu, Xiaolin and He, Xiaokai and Cao, Zhoujian",
    title = "{Accurate calculation of gravitational wave memory}",
    eprint = "2302.02642",
    archivePrefix = "arXiv",
    primaryClass = "gr-qc",
    doi = "10.1103/PhysRevD.103.043005",
    journal = "Phys. Rev. D",
    volume = "103",
    number = "4",
    pages = "043005",
    year = "2021"
}

@article{LISA:2024hlh,
	title        = {{LISA Definition Study Report}},
	author       = {Colpi, Monica and others},
	year         = 2024,
	month        = 2,
	collaboration = {LISA},
	eprint       = {2402.07571},
	archiveprefix = {arXiv},
	primaryclass = {astro-ph.CO}
}

@article{TianQin:2020hid,
	title        = {{The TianQin project: current progress on science and technology}},
	author       = {Mei, Jianwei and others},
	year         = 2021,
	journal      = {PTEP},
	volume       = 2021,
	number       = 5,
	pages        = {05A107},
	doi          = {10.1093/ptep/ptaa114},
	collaboration = {TianQin},
	eprint       = {2008.10332},
	archiveprefix = {arXiv},
	primaryclass = {gr-qc}
}

@article{Torii:1996yi,
	title        = {{Dilatonic black holes with Gauss-Bonnet term}},
	author       = {Torii, Takashi and Yajima, Hiroki and Maeda, Kei-ichi},
	year         = 1997,
	journal      = {Phys. Rev. D},
	volume       = 55,
	pages        = {739--753},
	doi          = {10.1103/PhysRevD.55.739},
	eprint       = {gr-qc/9606034},
	archiveprefix = {arXiv},
	reportnumber = {WU-AP-55-96}
}

@article{Tahura:2021hbk,
	title        = {{Gravitational-wave memory effects in Brans-Dicke theory: Waveforms and effects in the post-Newtonian approximation}},
	author       = {Tahura, Shammi and Nichols, David A. and Yagi, Kent},
	year         = 2021,
	journal      = {Phys. Rev. D},
	volume       = 104,
	number       = 10,
	pages        = 104010,
	doi          = {10.1103/PhysRevD.104.104010},
	eprint       = {2107.02208},
	archiveprefix = {arXiv},
	primaryclass = {gr-qc}
}

@article{ET:2025xjr,
	title        = {{The Science of the Einstein Telescope}},
	author       = {Abac, Adrian and others},
	year         = 2025,
	month        = 3,
	collaboration = {ET},
	eprint       = {2503.12263},
	archiveprefix = {arXiv},
	primaryclass = {gr-qc},
	reportnumber = {ET-0036C-25}
}

@article{Maibach:2026wpz,
    author = "Maibach, David and Zosso, Jann",
    title = "{Balance flux laws beyond general relativity}",
    eprint = "2601.07091",
    archivePrefix = "arXiv",
    primaryClass = "gr-qc",
    doi = "10.1103/8ndd-y8dk",
    journal = "Phys. Rev. D",
    volume = "113",
    number = "12",
    pages = "124085",
    year = "2026"
}

@article{Zosso:2026czc,
    author = "Zosso, Jann and Maga{\~n}a Zertuche, Lorena and Gasparotto, Silvia and Cogez, Adrien and Inchausp{\'e}, Henri and Jacobs, Milo",
    title = "{Toward claiming a detection of gravitational memory}",
    eprint = "2601.23019",
    archivePrefix = "arXiv",
    primaryClass = "gr-qc",
    doi = "10.1103/51xv-zlfy",
    journal = "Phys. Rev. D",
    volume = "113",
    number = "10",
    pages = "104033",
    year = "2026"
}

@inproceedings{Zosso:2025ffy,
	title        = {{Continuing Isaacson's Legacy: A general metric theory perspective on gravitational memory and the non-linearity of gravity}},
	author       = {Zosso, Jann},
	year         = 2025,
	month        = 5,
	booktitle    = {{59th Rencontres de Moriond on Gravitation}: {Moriond 2025 Gravitation}},
	eprint       = {2505.17603},
	archiveprefix = {arXiv},
	primaryclass = {gr-qc}
}

@article{Gasparotto:2026bru,
	title        = {{Gravitational Memory from Hairy Binary Black Hole Mergers}},
	author       = {Gasparotto, Silvia and Zosso, Jann and Arest{\'e} Sal{\'o}, Llibert and Doneva, Daniela D. and Yazadjiev, Stoytcho S.},
	year         = 2026,
	month        = 4,
	eprint       = {2604.09350},
	archiveprefix = {arXiv},
	primaryclass = {gr-qc},
	reportnumber = {CERN-TH-2026-086}
}

@article{Witek:2018dmd,
	title        = {{Black holes and binary mergers in scalar Gauss-Bonnet gravity: scalar field dynamics}},
	author       = {Witek, Helvi and Gualtieri, Leonardo and Pani, Paolo and Sotiriou, Thomas P.},
	year         = 2019,
	journal      = {Phys. Rev. D},
	volume       = 99,
	number       = 6,
	pages        = {064035},
	doi          = {10.1103/PhysRevD.99.064035},
	eprint       = {1810.05177},
	archiveprefix = {arXiv},
	primaryclass = {gr-qc}
}

@article{Ma:2023sok,
	title        = {{Numerical simulations of black hole-neutron star mergers in scalar-tensor gravity}},
	author       = {Ma, Sizheng and Varma, Vijay and Stein, Leo C. and Foucart, Francois and Duez, Matthew D. and Kidder, Lawrence E. and Pfeiffer, Harald P. and Scheel, Mark A.},
	year         = 2023,
	journal      = {Phys. Rev. D},
	volume       = 107,
	number       = 12,
	pages        = 124051,
	doi          = {10.1103/PhysRevD.107.124051},
	eprint       = {2304.11836},
	archiveprefix = {arXiv},
	primaryclass = {gr-qc}
}

@article{Reitze:2019iox,
	title        = {{Cosmic Explorer: The U.S. Contribution to Gravitational-Wave Astronomy beyond LIGO}},
	author       = {Reitze, David and others},
	year         = 2019,
	journal      = {Bull. Am. Astron. Soc.},
	volume       = 51,
	number       = 7,
	pages        = {035},
	eprint       = {1907.04833},
	archiveprefix = {arXiv},
	primaryclass = {astro-ph.IM},
	reportnumber = {LIGO-P1900316}
}

@article{LIGOScientific:2025slb,
	title        = {{GWTC-4.0: Updating the Gravitational-Wave Transient Catalog with Observations from the First Part of the Fourth LIGO-Virgo-KAGRA Observing Run}},
	author       = {Abac, A. G. and others},
	year         = 2025,
	month        = 8,
	collaboration = {LIGO Scientific, VIRGO, KAGRA},
	eprint       = {2508.18082},
	archiveprefix = {arXiv},
	primaryclass = {gr-qc},
	reportnumber = {LIGO-P2400386}
}

@article{Zeldovich:1974gvh,
	title        = {{Radiation of gravitational waves by a cluster of superdense stars}},
	author       = {Zel'dovich, Y. B. and Polnarev, A. G.},
	year         = 1974,
	journal      = {Sov. Astron.},
	volume       = 18,
	pages        = 17
}

@article{Braginsky:1985vlg,
	title        = {{Kinematic Resonance and Memory Effect in Free Mass Gravitational Antennas}},
	author       = {Braginsky, V. B. and Grishchuk, L. P.},
	year         = 1985,
	journal      = {Sov. Phys. JETP},
	volume       = 62,
	pages        = {427--430}
}

@article{Braginsky:1987kwo,
	title        = {{Gravitational-wave bursts with memory and experimental prospects}},
	author       = {Braginsky, Vladimir B. and Thorne, Kip S.},
	year         = 1987,
	journal      = {Nature},
	volume       = 327,
	pages        = {123--125},
	doi          = {10.1038/327123a0}
}

@book{poisson2014gravity,
	title        = {Gravity: Newtonian, Post-Newtonian, Relativistic},
	author       = {Poisson, Eric and Will, Clifford M.},
	year         = 2014,
	publisher    = {Cambridge University Press},
	doi          = {10.1017/CBO9781139507486},
	place        = {Cambridge}
}

@article{PhysRevD.45.520,
	title        = {Gravitational-wave bursts with memory: The Christodoulou effect},
	author       = {Thorne, Kip S.},
	year         = 1992,
	month        = {Jan},
	journal      = {Phys. Rev. D},
	publisher    = {American Physical Society},
	volume       = 45,
	pages        = {520--524},
	doi          = {10.1103/PhysRevD.45.520},
	url          = {https://link.aps.org/doi/10.1103/PhysRevD.45.520},
	issue        = 2,
	numpages     = {0}
}

@article{Christodoulou:1991cr,
	title        = {{Nonlinear nature of gravitation and gravitational wave experiments}},
	author       = {Christodoulou, D.},
	year         = 1991,
	journal      = {Phys. Rev. Lett.},
	volume       = 67,
	pages        = {1486--1489},
	doi          = {10.1103/PhysRevLett.67.1486}
}

@article{Blanchet:1992br,
	title        = {{Hereditary effects in gravitational radiation}},
	author       = {Blanchet, Luc and Damour, Thibault},
	year         = 1992,
	journal      = {Phys. Rev. D},
	volume       = 46,
	pages        = {4304--4319},
	doi          = {10.1103/PhysRevD.46.4304},
	reportnumber = {IHES-P-92-42}
}

@article{Wiseman:1991ss,
	title        = {{Christodoulou's nonlinear gravitational wave memory: Evaluation in the quadrupole approximation}},
	author       = {Wiseman, Alan G. and Will, Clifford M.},
	year         = 1991,
	journal      = {Phys. Rev. D},
	volume       = 44,
	number       = 10,
	pages        = {R2945--R2949},
	doi          = {10.1103/PhysRevD.44.R2945},
	reportnumber = {WUGRAV-91-3}
}

@book{Strominger:2017zoo,
	title        = {{Lectures on the Infrared Structure of Gravity and Gauge Theory}},
	author       = {Strominger, Andrew},
	year         = 2018,
	publisher    = {Princeton University Press},
	isbn         = {978-0-691-17973-5},
	eprint       = {1703.05448},
	archiveprefix = {arXiv},
	primaryclass = {hep-th}
}

@article{Bondi:1962px,
	title        = {{Gravitational waves in general relativity. 7. Waves from axisymmetric isolated systems}},
	author       = {Bondi, H. and van der Burg, M. G. J. and Metzner, A. W. K.},
	year         = 1962,
	journal      = {Proc. Roy. Soc. Lond. A},
	volume       = 269,
	pages        = {21--52},
	doi          = {10.1098/rspa.1962.0161}
}

@article{Sachs:1962wk,
	title        = {{Gravitational waves in general relativity. 8. Waves in asymptotically flat space-times}},
	author       = {Sachs, R. K.},
	year         = 1962,
	journal      = {Proc. Roy. Soc. Lond. A},
	volume       = 270,
	pages        = {103--126},
	doi          = {10.1098/rspa.1962.0206}
}

@article{Strominger:2013jfa,
	title        = {{On BMS Invariance of Gravitational Scattering}},
	author       = {Strominger, Andrew},
	year         = 2014,
	journal      = {JHEP},
	volume       = {07},
	pages        = 152,
	doi          = {10.1007/JHEP07(2014)152},
	eprint       = {1312.2229},
	archiveprefix = {arXiv},
	primaryclass = {hep-th}
}

@article{He:2014laa,
	title        = {{BMS supertranslations and Weinberg{\textquoteright}s soft graviton theorem}},
	author       = {He, Temple and Lysov, Vyacheslav and Mitra, Prahar and Strominger, Andrew},
	year         = 2015,
	journal      = {JHEP},
	volume       = {05},
	pages        = 151,
	doi          = {10.1007/JHEP05(2015)151},
	eprint       = {1401.7026},
	archiveprefix = {arXiv},
	primaryclass = {hep-th}
}

@article{Strominger:2014pwa,
	title        = {{Gravitational Memory, BMS Supertranslations and Soft Theorems}},
	author       = {Strominger, Andrew and Zhiboedov, Alexander},
	year         = 2016,
	journal      = {JHEP},
	volume       = {01},
	pages        = {086},
	doi          = {10.1007/JHEP01(2016)086},
	eprint       = {1411.5745},
	archiveprefix = {arXiv},
	primaryclass = {hep-th}
}

@article{Pasterski:2015tva,
	title        = {{New Gravitational Memories}},
	author       = {Pasterski, Sabrina and Strominger, Andrew and Zhiboedov, Alexander},
	year         = 2016,
	journal      = {JHEP},
	volume       = 12,
	pages        = {053},
	doi          = {10.1007/JHEP12(2016)053},
	eprint       = {1502.06120},
	archiveprefix = {arXiv},
	primaryclass = {hep-th}
}

@article{Favata:2009ii,
	title        = {{Nonlinear gravitational-wave memory from binary black hole mergers}},
	author       = {Favata, Marc},
	year         = 2009,
	journal      = {Astrophys. J. Lett.},
	volume       = 696,
	pages        = {L159--L162},
	doi          = {10.1088/0004-637X/696/2/L159},
	eprint       = {0902.3660},
	archiveprefix = {arXiv},
	primaryclass = {astro-ph.SR}
}

@article{Tahura:2020vsa,
	title        = {{Brans-Dicke theory in Bondi-Sachs form: Asymptotically flat solutions, asymptotic symmetries and gravitational-wave memory effects}},
	author       = {Tahura, Shammi and Nichols, David A. and Saffer, Alexander and Stein, Leo C. and Yagi, Kent},
	year         = 2021,
	journal      = {Phys. Rev. D},
	volume       = 103,
	number       = 10,
	pages        = 104026,
	doi          = {10.1103/PhysRevD.103.104026},
	eprint       = {2007.13799},
	archiveprefix = {arXiv},
	primaryclass = {gr-qc}
}

@article{Hou:2020tnd,
	title        = {{Gravitational memory effects and Bondi-Metzner-Sachs symmetries in scalar-tensor theories}},
	author       = {Hou, Shaoqi and Zhu, Zong-Hong},
	year         = 2021,
	journal      = {JHEP},
	volume       = {01},
	pages        = {083},
	doi          = {10.1007/JHEP01(2021)083},
	eprint       = {2005.01310},
	archiveprefix = {arXiv},
	primaryclass = {gr-qc}
}

@article{Hou:2020wbo,
	title        = {{''Conserved charges'' of the Bondi-Metzner-Sachs algebra in the Brans-Dicke theory}},
	author       = {Hou, Shaoqi and Zhu, Zong-Hong},
	year         = 2021,
	journal      = {Chin. Phys. C},
	volume       = 45,
	number       = 2,
	pages        = {023122},
	doi          = {10.1088/1674-1137/abd087},
	eprint       = {2008.05154},
	archiveprefix = {arXiv},
	primaryclass = {gr-qc}
}

@article{Tahura:2025ebb,
	title        = {{Gravitational-wave memory effects in the Damour-Esposito-Far{\`e}se extension of Brans-Dicke theory}},
	author       = {Tahura, Shammi and Nichols, David A. and Yagi, Kent},
	year         = 2025,
	journal      = {Phys. Rev. D},
	volume       = 112,
	number       = 8,
	pages        = {084037},
	doi          = {10.1103/c657-4dd5},
	eprint       = {2501.07488},
	archiveprefix = {arXiv},
	primaryclass = {gr-qc}
}

@article{Lang:2013fna,
	title        = {{Compact binary systems in scalar-tensor gravity. II. Tensor gravitational waves to second post-Newtonian order}},
	author       = {Lang, Ryan N.},
	year         = 2014,
	journal      = {Phys. Rev. D},
	volume       = 89,
	number       = 8,
	pages        = {084014},
	doi          = {10.1103/PhysRevD.89.084014},
	eprint       = {1310.3320},
	archiveprefix = {arXiv},
	primaryclass = {gr-qc}
}

@article{Du:2016hww,
	title        = {{Gravitational Wave Memory: A New Approach to Study Modified Gravity}},
	author       = {Du, Song Ming and Nishizawa, Atsushi},
	year         = 2016,
	journal      = {Phys. Rev. D},
	volume       = 94,
	number       = 10,
	pages        = 104063,
	doi          = {10.1103/PhysRevD.94.104063},
	eprint       = {1609.09825},
	archiveprefix = {arXiv},
	primaryclass = {gr-qc}
}

@article{Heisenberg:2025tfh,
	title        = {{Constraining superluminal Einstein-{\AE}ther gravity through gravitational memory}},
	author       = {Heisenberg, Lavinia and Rosatello, Benedetta and Xu, Guangzi and Zosso, Jann},
	year         = 2025,
	journal      = {Phys. Rev. D},
	volume       = 112,
	number       = 2,
	pages        = {024052},
	doi          = {10.1103/2zds-qq93},
	eprint       = {2505.09544},
	archiveprefix = {arXiv},
	primaryclass = {gr-qc}
}

@article{Kovacs:2020ywu,
	title        = {{Well-posed formulation of Lovelock and Horndeski theories}},
	author       = {Kov{\'a}cs, {\'A}ron D. and Reall, Harvey S.},
	year         = 2020,
	journal      = {Phys. Rev. D},
	volume       = 101,
	number       = 12,
	pages        = 124003,
	doi          = {10.1103/PhysRevD.101.124003},
	eprint       = {2003.08398},
	archiveprefix = {arXiv},
	primaryclass = {gr-qc}
}

@article{Kovacs:2020pns,
	title        = {{Well-Posed Formulation of Scalar-Tensor Effective Field Theory}},
	author       = {Kov{\'a}cs, {\'A}ron D. and Reall, Harvey S.},
	year         = 2020,
	journal      = {Phys. Rev. Lett.},
	volume       = 124,
	number       = 22,
	pages        = 221101,
	doi          = {10.1103/PhysRevLett.124.221101},
	eprint       = {2003.04327},
	archiveprefix = {arXiv},
	primaryclass = {gr-qc}
}

@article{Koyama:2020vfc,
	title        = {{Testing Brans-Dicke Gravity with Screening by Scalar Gravitational Wave Memory}},
	author       = {Koyama, Kazuya},
	year         = 2020,
	journal      = {Phys. Rev. D},
	volume       = 102,
	number       = 2,
	pages        = {021502},
	doi          = {10.1103/PhysRevD.102.021502},
	eprint       = {2006.15914},
	archiveprefix = {arXiv},
	primaryclass = {gr-qc}
}

@article{Heisenberg:2023prj,
	title        = {{Gravitational wave memory beyond general relativity}},
	author       = {Heisenberg, Lavinia and Yunes, Nicol\'as and Zosso, Jann},
	year         = 2023,
	journal      = {Phys. Rev. D},
	volume       = 108,
	number       = 2,
	pages        = {024010},
	doi          = {10.1103/PhysRevD.108.024010},
	eprint       = {2303.02021},
	archiveprefix = {arXiv},
	primaryclass = {gr-qc}
}

@article{Bertotti:2003rm,
	title        = {{A test of general relativity using radio links with the Cassini spacecraft}},
	author       = {Bertotti, B. and Iess, L. and Tortora, P.},
	year         = 2003,
	journal      = {Nature},
	volume       = 425,
	pages        = {374--376},
	doi          = {10.1038/nature01997}
}

@article{Reasenberg:1979ey,
	title        = {{Viking relativity experiment: Verification of signal retardation by solar gravity}},
	author       = {Reasenberg, R. D. and Shapiro, I. I. and MacNeil, P. E. and Goldstein, R. B. and Breidenthal, J. C. and Brenkle, J. P. and Cain, D. L. and Kaufman, T. M. and Komarek, T. A. and Zygielbaum, A. I.},
	year         = 1979,
	journal      = {Astrophys. J. Lett.},
	volume       = 234,
	pages        = {L219--L221},
	doi          = {10.1086/183144}
}

@article{Will:2014kxa,
	title        = {{The Confrontation between General Relativity and Experiment}},
	author       = {Will, Clifford M.},
	year         = 2014,
	journal      = {Living Rev. Rel.},
	volume       = 17,
	pages        = 4,
	doi          = {10.12942/lrr-2014-4},
	eprint       = {1403.7377},
	archiveprefix = {arXiv},
	primaryclass = {gr-qc}
}

@article{Perkins:2021mhb,
	title        = {{Improved gravitational-wave constraints on higher-order curvature theories of gravity}},
	author       = {Perkins, Scott E. and Nair, Remya and Silva, Hector O. and Yunes, Nicolas},
	year         = 2021,
	journal      = {Phys. Rev. D},
	volume       = 104,
	number       = 2,
	pages        = {024060},
	doi          = {10.1103/PhysRevD.104.024060},
	eprint       = {2104.11189},
	archiveprefix = {arXiv},
	primaryclass = {gr-qc}
}

@article{Yordanov:2024lfk,
    author = "Yordanov, Petar Y. and Staykov, Kalin V. and Yazadjiev, Stoytcho S. and Doneva, Daniela D.",
    title = "{The power of binary pulsars in testing Gauss-Bonnet gravity}",
    eprint = "2402.06305",
    archivePrefix = "arXiv",
    primaryClass = "gr-qc",
    doi = "10.1051/0004-6361/202449679",
    journal = "Astron. Astrophys.",
    volume = "687",
    pages = "A17",
    year = "2024"
}

@article{Lyu:2022gdr,
	title        = {{Constraints on Einstein-dilation-Gauss-Bonnet gravity from black hole-neutron star gravitational wave events}},
	author       = {Lyu, Zhenwei and Jiang, Nan and Yagi, Kent},
	year         = 2022,
	journal      = {Phys. Rev. D},
	volume       = 105,
	number       = 6,
	pages        = {064001},
	doi          = {10.1103/PhysRevD.105.064001},
	note         = {[Erratum: Phys.Rev.D 106, 069901 (2022), Erratum: Phys.Rev.D 106, 069901 (2022)]},
	eprint       = {2201.02543},
	archiveprefix = {arXiv},
	primaryclass = {gr-qc},
	reportnumber = {LIGO-P2100466}
}

@article{Wang:2023wgv,
	title        = {{Constraining the Einstein-dilaton-Gauss-Bonnet theory with higher harmonics and the merger-ringdown contribution using GWTC-3}},
	author       = {Wang, Baoxiang and Shi, Changfu and Zhang, Jian-dong and hu, Yi-Ming and Mei, Jianwei},
	year         = 2023,
	journal      = {Phys. Rev. D},
	volume       = 108,
	number       = 4,
	pages        = {044061},
	doi          = {10.1103/PhysRevD.108.044061},
	eprint       = {2302.10112},
	archiveprefix = {arXiv},
	primaryclass = {gr-qc}
}

@article{PhysRevLett.92.121101,
	title        = {Measurement of the Solar Gravitational Deflection of Radio Waves using Geodetic Very-Long-Baseline Interferometry Data, 1979--1999},
	author       = {Shapiro, S. S. and Davis, J. L. and Lebach, D. E. and Gregory, J. S.},
	year         = 2004,
	month        = {Mar},
	journal      = {Phys. Rev. Lett.},
	publisher    = {American Physical Society},
	volume       = 92,
	pages        = 121101,
	doi          = {10.1103/PhysRevLett.92.121101},
	issue        = 12,
	numpages     = 4
}

@article{Flanagan:2005yc,
	title        = {The Basics of gravitational wave theory},
	author       = {Flanagan, Eanna E. and Hughes, Scott A.},
	year         = 2005,
	journal      = {New J. Phys.},
	volume       = 7,
	pages        = 204,
	doi          = {10.1088/1367-2630/7/1/204},
	eprint       = {gr-qc/0501041},
	archiveprefix = {arXiv}
}

@article{Moura:2006pz,
	title        = {{Higher-derivative corrected black holes: Perturbative stability and absorption cross-section in heterotic string theory}},
	author       = {Moura, Filipe and Schiappa, Ricardo},
	year         = 2007,
	journal      = {Class. Quant. Grav.},
	volume       = 24,
	pages        = {361--386},
	doi          = {10.1088/0264-9381/24/2/006},
	eprint       = {hep-th/0605001},
	archiveprefix = {arXiv},
	reportnumber = {CPHT-RR047-0805, SPHT-T05-51, ITFA-2006-19, CERN-PH-TH-2006-076}
}

@article{Doneva:2022ewd,
	title        = {{Spontaneous scalarization}},
	author       = {Doneva, Daniela D. and Ramazano\u{g}lu, Fethi M. and Silva, Hector O. and Sotiriou, Thomas P. and Yazadjiev, Stoytcho S.},
	year         = 2024,
	journal      = {Rev. Mod. Phys.},
	volume       = 96,
	number       = 1,
	pages        = {015004},
	doi          = {10.1103/RevModPhys.96.015004},
	eprint       = {2211.01766},
	archiveprefix = {arXiv},
	primaryclass = {gr-qc}
}

@article{Heisenberg:2024cjk,
	title        = {{Unifying ordinary and null memory}},
	author       = {Heisenberg, Lavinia and Xu, Guangzi and Zosso, Jann},
	year         = 2024,
	journal      = {JCAP},
	volume       = {05},
	pages        = 119,
	doi          = {10.1088/1475-7516/2024/05/119},
	eprint       = {2401.05936},
	archiveprefix = {arXiv},
	primaryclass = {gr-qc}
}

@article{Doneva:2024ntw,
    author = "Doneva, Daniela D. and Arest{\'e} Sal{\'o}, Llibert and Yazadjiev, Stoytcho S.",
    title = "{3+1 nonlinear evolution of Ricci-coupled scalar-Gauss-Bonnet gravity}",
    eprint = "2404.15526",
    archivePrefix = "arXiv",
    primaryClass = "gr-qc",
    doi = "10.1103/PhysRevD.110.024040",
    journal = "Phys. Rev. D",
    volume = "110",
    number = "2",
    pages = "024040",
    year = "2024"
}

@article{Julie:2023ncq,
	title        = {{Dynamical scalarization in Schwarzschild binary inspirals}},
	author       = {Juli{\'e}, F{\'e}lix-Louis},
	year         = 2023,
	month        = 12,
	eprint       = {2312.16764},
	archiveprefix = {arXiv},
	primaryclass = {gr-qc}
}

@article{Evstafyeva:2022rve,
	title        = {{Measuring the ringdown scalar polarization of gravitational waves in Einstein-scalar-Gauss-Bonnet gravity}},
	author       = {Evstafyeva, Tamara and Agathos, Michalis and Ripley, Justin L.},
	year         = 2023,
	journal      = {Phys. Rev. D},
	volume       = 107,
	number       = 12,
	pages        = 124010,
	doi          = {10.1103/PhysRevD.107.124010},
	eprint       = {2212.11359},
	archiveprefix = {arXiv},
	primaryclass = {gr-qc}
}

@article{Metsaev:1987zx,
	title        = {{Order alpha-prime (Two Loop) Equivalence of the String Equations of Motion and the Sigma Model Weyl Invariance Conditions: Dependence on the Dilaton and the Antisymmetric Tensor}},
	author       = {Metsaev, R. R. and Tseytlin, Arkady A.},
	year         = 1987,
	journal      = {Nucl. Phys. B},
	volume       = 293,
	pages        = {385--419},
	doi          = {10.1016/0550-3213(87)90077-0},
	reportnumber = {PRINT-87-0184 (LEBEDEV)}
}

@phdthesis{Zosso:2024xgy,
	title        = {{Probing Gravity - Fundamental Aspects of Metric Theories and their Implications for Tests of General Relativity}},
	author       = {Zosso, Jann},
	year         = 2024,
	doi          = {10.3929/ethz-b-000675938},
	eprint       = {2412.06043},
	archiveprefix = {arXiv},
	primaryclass = {gr-qc},
	school       = {Zurich, ETH}
}

@article{Zwiebach:1985uq,
	title        = {Curvature Squared Terms and String Theories},
	author       = {Zwiebach, Barton},
	year         = 1985,
	journal      = {Phys. Lett. B},
	volume       = 156,
	pages        = {315--317},
	doi          = {10.1016/0370-2693(85)91616-8},
	reportnumber = {UCB-PTH-85/10}
}

@article{Boulware:1986dr,
	title        = {{Effective Gravity Theories With Dilatons}},
	author       = {Boulware, David G. and Deser, Stanley},
	year         = 1986,
	journal      = {Phys. Lett. B},
	volume       = 175,
	pages        = {409--412},
	doi          = {10.1016/0370-2693(86)90614-3},
	reportnumber = {NSF-ITP-86-57}
}

@article{Nojiri:2017ncd,
	title        = {Modified Gravity Theories on a Nutshell: Inflation, Bounce and Late-time Evolution},
	author       = {Nojiri, S. and Odintsov, S. D. and Oikonomou, V. K.},
	year         = 2017,
	journal      = {Phys. Rept.},
	volume       = 692,
	pages        = {1--104},
	doi          = {10.1016/j.physrep.2017.06.001},
	eprint       = {1705.11098},
	archiveprefix = {arXiv},
	primaryclass = {gr-qc},
	reportnumber = {PHYS.REPT.-692-(2017)-1-104, Phys.Rept. 692 (2017) 1-104}
}

@article{Ortega:2024prv,
	title        = {{Field Equations in Chern-Simons-Gauss-Bonnet Gravity}},
	author       = {Ortega, Alexis and Daniel, Tatsuya and Koushiappas, Savvas M.},
	year         = 2024,
	month        = 11,
	eprint       = {2411.05911},
	archiveprefix = {arXiv},
	primaryclass = {gr-qc},
	reportnumber = {MIT-CTP/5806}
}

@article{Weinberg:2008hq,
	title        = {Effective Field Theory for Inflation},
	author       = {Weinberg, Steven},
	year         = 2008,
	journal      = {Phys. Rev. D},
	volume       = 77,
	pages        = 123541,
	doi          = {10.1103/PhysRevD.77.123541},
	eprint       = {0804.4291},
	archiveprefix = {arXiv},
	primaryclass = {hep-th},
	reportnumber = {UTTG-01-08}
}

@book{misner_gravitation_1973,
	title        = {Gravitation},
	author       = {Misner, Charles W. and Thorne, K. S. and Wheeler, J. A.},
	year         = 1973,
	publisher    = {W. H. Freeman},
	address      = {San Francisco},
	isbn         = {978-0-7167-0344-0, 978-0-691-17779-3}
}

@book{maggiore2008gravitational,
	title        = {Gravitational Waves: Volume 1: Theory and Experiments},
	author       = {Maggiore, Michele},
	year         = 2007,
	month        = 10,
	publisher    = {Oxford University Press},
	doi          = {10.1093/acprof:oso/9780198570745.001.0001},
	isbn         = 9780198570745
}

@book{carroll2019spacetime,
	title        = {Spacetime and Geometry: An Introduction to General Relativity},
	author       = {Carroll, Sean M.},
	year         = 2019,
	publisher    = {Cambridge University Press},
	doi          = {10.1017/9781108770385},
	place        = {Cambridge}
}

@article{Gross:1986iv,
	title        = {Superstring Modifications of Einstein's Equations},
	author       = {Gross, David J. and Witten, Edward},
	year         = 1986,
	journal      = {Nucl. Phys. B},
	volume       = 277,
	pages        = 1,
	doi          = {10.1016/0550-3213(86)90429-3},
	reportnumber = {Print-86-0250 (PRINCETON)}
}

@article{Jackiw:2003pm,
	title        = {Chern-Simons modification of general relativity},
	author       = {Jackiw, R. and Pi, S. Y.},
	year         = 2003,
	journal      = {Phys. Rev. D},
	volume       = 68,
	pages        = 104012,
	doi          = {10.1103/PhysRevD.68.104012},
	eprint       = {gr-qc/0308071},
	archiveprefix = {arXiv},
	reportnumber = {MIT-CTP-3409, BUHEP-03-18}
}

@article{Simon:1990PhysRevD41,
	title        = {Higher-derivative Lagrangians, nonlocality, problems, and solutions},
	author       = {Simon, Jonathan Z.},
	year         = 1990,
	month        = {Jun},
	journal      = {Phys. Rev. D},
	publisher    = {American Physical Society},
	volume       = 41,
	pages        = {3720--3733},
	doi          = {10.1103/PhysRevD.41.3720},
	issue        = 12,
	numpages     = {0}
}

@article{Yunes:2013dva,
	title        = {Gravitational-Wave Tests of General Relativity with Ground-Based Detectors and Pulsar Timing-Arrays},
	author       = {Yunes, Nicol\'as and Siemens, Xavier},
	year         = 2013,
	journal      = {Living Rev. Rel.},
	volume       = 16,
	pages        = 9,
	doi          = {10.12942/lrr-2013-9},
	eprint       = {1304.3473},
	archiveprefix = {arXiv},
	primaryclass = {gr-qc}
}

@article{Silva:2020omi,
	title        = {Dynamical Descalarization in Binary Black Hole Mergers},
	author       = {Silva, Hector O. and Witek, Helvi and Elley, Matthew and Yunes, Nicol\'as},
	year         = 2021,
	journal      = {Phys. Rev. Lett.},
	volume       = 127,
	number       = 3,
	pages        = {031101},
	doi          = {10.1103/PhysRevLett.127.031101},
	eprint       = {2012.10436},
	archiveprefix = {arXiv},
	primaryclass = {gr-qc}
}

@article{Elley:2022ept,
	title        = {Spin-induced dynamical scalarization, descalarization, and stealthness in scalar-Gauss-Bonnet gravity during a black hole coalescence},
	author       = {Elley, Matthew and Silva, Hector O. and Witek, Helvi and Yunes, Nicol\'as},
	year         = 2022,
	journal      = {Phys. Rev. D},
	volume       = 106,
	number       = 4,
	pages        = {044018},
	doi          = {10.1103/PhysRevD.106.044018},
	eprint       = {2205.06240},
	archiveprefix = {arXiv},
	primaryclass = {gr-qc}
}

@article{Heisenberg:2018vsk,
	title        = {A systematic approach to generalisations of General Relativity and their cosmological implications},
	author       = {Heisenberg, Lavinia},
	year         = 2019,
	journal      = {Phys. Rept.},
	volume       = 796,
	pages        = {1--113},
	doi          = {10.1016/j.physrep.2018.11.006},
	eprint       = {1807.01725},
	archiveprefix = {arXiv},
	primaryclass = {gr-qc},
	slaccitation = {%%CITATION = ARXIV:1807.01725;%%}
}

@article{Nicolis:2008in,
	title        = {The Galileon as a local modification of gravity},
	author       = {Nicolis, Alberto and Rattazzi, Riccardo and Trincherini, Enrico},
	year         = 2009,
	journal      = {Phys. Rev. D},
	volume       = 79,
	pages        = {064036},
	doi          = {10.1103/PhysRevD.79.064036},
	eprint       = {0811.2197},
	archiveprefix = {arXiv},
	primaryclass = {hep-th},
	slaccitation = {%%CITATION = ARXIV:0811.2197;%%}
}

@article{Deffayet:2009wt,
	title        = {Covariant Galileon},
	author       = {Deffayet, C. and Esposito-Farese, Gilles and Vikman, A.},
	year         = 2009,
	journal      = {Phys. Rev. D},
	volume       = 79,
	pages        = {084003},
	doi          = {10.1103/PhysRevD.79.084003},
	eprint       = {0901.1314},
	archiveprefix = {arXiv},
	primaryclass = {hep-th},
	slaccitation = {%%CITATION = ARXIV:0901.1314;%%}
}

@article{Deffayet:2009mn,
	title        = {Generalized Galileons: All scalar models whose curved background extensions maintain second-order field equations and stress-tensors},
	author       = {Deffayet, C. and Deser, S. and Esposito-Farese, G.},
	year         = 2009,
	journal      = {Phys. Rev. D},
	volume       = 80,
	pages        = {064015},
	doi          = {10.1103/PhysRevD.80.064015},
	eprint       = {0906.1967},
	archiveprefix = {arXiv},
	primaryclass = {gr-qc},
	slaccitation = {%%CITATION = ARXIV:0906.1967;%%}
}

@article{Horndeski:1974wa,
	title        = {Second-order scalar-tensor field equations in a four-dimensional space},
	author       = {Horndeski, Gregory Walter},
	year         = 1974,
	journal      = {Int. J. Theor. Phys.},
	volume       = 10,
	pages        = {363--384},
	doi          = {10.1007/BF01807638},
	slaccitation = {%%CITATION = IJTPB,10,363;%%}
}

@article{Kobayashi:2019hrl,
	title        = {Horndeski theory and beyond: a review},
	author       = {Kobayashi, Tsutomu},
	year         = 2019,
	journal      = {Rept. Prog. Phys.},
	volume       = 82,
	number       = 8,
	pages        = {086901},
	doi          = {10.1088/1361-6633/ab2429},
	eprint       = {1901.07183},
	archiveprefix = {arXiv},
	primaryclass = {gr-qc},
	reportnumber = {RUP-19-3}
}

@article{Hou:2017bqj,
	title        = {Polarizations of Gravitational Waves in Horndeski Theory},
	author       = {Hou, Shaoqi and Gong, Yungui and Liu, Yunqi},
	year         = 2018,
	journal      = {Eur. Phys. J. C},
	volume       = 78,
	number       = 5,
	pages        = 378,
	doi          = {10.1140/epjc/s10052-018-5869-y},
	eprint       = {1704.01899},
	archiveprefix = {arXiv},
	primaryclass = {gr-qc}
}

@article{Ripley:2022cdh,
	title        = {Numerical relativity for Horndeski gravity},
	author       = {Ripley, Justin L.},
	year         = 2022,
	journal      = {Int. J. Mod. Phys. D},
	volume       = 31,
	number       = 13,
	pages        = 2230017,
	doi          = {10.1142/S0218271822300178},
	eprint       = {2207.13074},
	archiveprefix = {arXiv},
	primaryclass = {gr-qc}
}

@article{Alexander:2009tp,
	title        = {{Chern-Simons Modified General Relativity}},
	author       = {Alexander, Stephon and Yunes, Nicolas},
	year         = 2009,
	journal      = {Phys. Rept.},
	volume       = 480,
	pages        = {1--55},
	doi          = {10.1016/j.physrep.2009.07.002},
	eprint       = {0907.2562},
	archiveprefix = {arXiv},
	primaryclass = {hep-th}
}

@book{Will:2018bme,
	title        = {Theory and Experiment in Gravitational Physics},
	author       = {Will, Clifford M.},
	year         = 2018,
	publisher    = {Cambridge University Press},
	doi          = {10.1017/9781316338612},
	place        = {Cambridge},
	edition      = 2
}

@incollection{papantonopoulos2014modifications,
	title        = {Gravity and Scalar Fields},
	author       = {Sotiriou, Thomas P.},
	year         = 2015,
	booktitle    = {Modifications of Einstein's Theory of Gravity at Large Distances},
	publisher    = {Springer International Publishing},
	address      = {Cham},
	pages        = {3--24},
	doi          = {10.1007/978-3-319-10070-8_1},
	editor       = {Papantonopoulos, Eleftherios}
}

@article{Seraj:2021qja,
	title        = {{Gravitational breathing memory and dual symmetries}},
	author       = {Seraj, Ali},
	year         = 2021,
	journal      = {JHEP},
	volume       = {05},
	pages        = 283,
	doi          = {10.1007/JHEP05(2021)283},
	eprint       = {2103.12185},
	archiveprefix = {arXiv},
	primaryclass = {hep-th}
}

@article{Corman:2025wun,
	title        = {{Black hole binaries in shift-symmetric Einstein-scalar-Gauss-Bonnet gravity experience a slower merger phase}},
	author       = {Corman, Maxence and Arest{\'e} Sal{\'o}, Llibert and Clough, Katy},
	year         = 2025,
	month        = 11,
	journal      = {},
	eprint       = {2511.19073},
	archiveprefix = {arXiv},
	primaryclass = {gr-qc}
}

@article{AresteSalo:2025sxc,
	title        = {{Challenges in the nonlinear evolution of unequal mass binaries in scalar-Gauss-Bonnet gravity}},
	author       = {Arest{\'e} Sal{\'o}, Llibert and Doneva, Daniela D. and Clough, Katy and Figueras, Pau and Yazadjiev, Stoytcho S.},
	year         = 2025,
	journal      = {Phys. Rev. D},
	volume       = 112,
	number       = 8,
	pages        = {084022},
	doi          = {10.1103/tr7v-jhhm},
	eprint       = {2507.13046},
	archiveprefix = {arXiv},
	primaryclass = {gr-qc}
}

@article{AresteSalo:2023hcp,
	title        = {{GRFolres: A code for modified gravity simulations in strong gravity}},
	author       = {Arest{\'e} Sal{\'o}, Llibert and Brady, Sam E. and Clough, Katy and Doneva, Daniela and Evstafyeva, Tamara and Figueras, Pau and Fran{\c{c}}a, Tiago and Rossi, Lorenzo and Yao, Shunhui},
	year         = 2024,
	journal      = {J. Open Source Softw.},
	volume       = 9,
	number       = 98,
	pages        = 6369,
	doi          = {10.21105/joss.06369},
	eprint       = {2309.06225},
	archiveprefix = {arXiv},
	primaryclass = {gr-qc}
}

@article{Andrade:2021rbd,
	title        = {{GRChombo: An adaptable numerical relativity code for fundamental physics}},
	author       = {Andrade, Tomas and others},
	year         = 2021,
	journal      = {J. Open Source Softw.},
	volume       = 6,
	number       = 68,
	pages        = 3703,
	doi          = {10.21105/joss.03703},
	eprint       = {2201.03458},
	archiveprefix = {arXiv},
	primaryclass = {gr-qc}
}

@article{Capuano:2026lhs,
	title        = {{Dynamical hair growth in black hole binaries in Einstein-scalar-Gauss-Bonnet gravity}},
	author       = {Capuano, Lodovico and Arest{\'e} Sal{\'o}, Llibert and Doneva, Daniela D. and Yazadjiev, Stoytcho S. and Barausse, Enrico},
	year         = 2026,
	month        = 2,
	eprint       = {2602.02650},
	archiveprefix = {arXiv},
	primaryclass = {gr-qc}
}

@article{Isaacson_PhysRev.166.1272,
	title        = {Gravitational Radiation in the Limit of High Frequency. II. Nonlinear Terms and the Effective Stress Tensor},
	author       = {Isaacson, Richard A.},
	year         = 1968,
	month        = {Feb},
	journal      = {Phys. Rev.},
	publisher    = {American Physical Society},
	volume       = 166,
	pages        = {1272--1280},
	doi          = {10.1103/PhysRev.166.1272},
	issue        = 5,
	numpages     = {0}
}

@article{Isaacson_PhysRev.166.1263,
	title        = {Gravitational Radiation in the Limit of High Frequency. I. The Linear Approximation and Geometrical Optics},
	author       = {Isaacson, Richard A.},
	year         = 1968,
	month        = {Feb},
	journal      = {Phys. Rev.},
	publisher    = {American Physical Society},
	volume       = 166,
	pages        = {1263--1271},
	doi          = {10.1103/PhysRev.166.1263},
	issue        = 5,
	numpages     = {0}
}

@article{Weinberg_PhysRev.140.B516,
	title        = {Infrared Photons and Gravitons},
	author       = {Weinberg, Steven},
	year         = 1965,
	month        = {Oct},
	journal      = {Phys. Rev.},
	publisher    = {American Physical Society},
	volume       = 140,
	pages        = {B516--B524},
	doi          = {10.1103/PhysRev.140.B516},
	issue        = {2B},
	numpages     = {0}
}

@article{Heisenberg:2025roe,
	title        = {{Gravitational memory in generalized Proca gravity}},
	author       = {Heisenberg, Lavinia and Rosatello, Benedetta and Xu, Guangzi and Zosso, Jann},
	year         = 2025,
	journal      = {Phys. Rev. D},
	volume       = 112,
	number       = 10,
	pages        = 104073,
	doi          = {10.1103/k24l-h7yy},
	eprint       = {2508.20545},
	archiveprefix = {arXiv},
	primaryclass = {gr-qc}
}

@article{Goncharov:2023woe,
	title        = {{Inferring Fundamental Spacetime Symmetries with Gravitational-Wave Memory: From LISA to the Einstein Telescope}},
	author       = {Goncharov, Boris and Donnay, Laura and Harms, Jan},
	year         = 2024,
	journal      = {Phys. Rev. Lett.},
	volume       = 132,
	number       = 24,
	pages        = 241401,
	doi          = {10.1103/PhysRevLett.132.241401},
	eprint       = {2310.10718},
	archiveprefix = {arXiv},
	primaryclass = {gr-qc}
}

@article{Hubner:2019sly,
	title        = {{Measuring gravitational-wave memory in the first LIGO/Virgo gravitational-wave transient catalog}},
	author       = {H{\"u}bner, Moritz and Talbot, Colm and Lasky, Paul D. and Thrane, Eric},
	year         = 2020,
	journal      = {Phys. Rev. D},
	volume       = 101,
	number       = 2,
	pages        = {023011},
	doi          = {10.1103/PhysRevD.101.023011},
	eprint       = {1911.12496},
	archiveprefix = {arXiv},
	primaryclass = {astro-ph.HE}
}

@article{Hubner:2021amk,
	title        = {{Memory remains undetected: Updates from the second LIGO/Virgo gravitational-wave transient catalog}},
	author       = {H{\"u}bner, Moritz and Lasky, Paul and Thrane, Eric},
	year         = 2021,
	journal      = {Phys. Rev. D},
	volume       = 104,
	number       = 2,
	pages        = {023004},
	doi          = {10.1103/PhysRevD.104.023004},
	eprint       = {2105.02879},
	archiveprefix = {arXiv},
	primaryclass = {gr-qc}
}

@article{Grant:2022bla,
	title        = {{Outlook for detecting the gravitational-wave displacement and spin memory effects with current and future gravitational-wave detectors}},
	author       = {Grant, Alexander M. and Nichols, David A.},
	year         = 2023,
	journal      = {Phys. Rev. D},
	volume       = 107,
	number       = 6,
	pages        = {064056},
	doi          = {10.1103/PhysRevD.107.064056},
	note         = {[Erratum: Phys.Rev.D 108, 029901 (2023)]},
	eprint       = {2210.16266},
	archiveprefix = {arXiv},
	primaryclass = {gr-qc}
}

@article{Gasparotto:2023fcg,
	title        = {{Can gravitational-wave memory help constrain binary black-hole parameters? A LISA case study}},
	author       = {Gasparotto, Silvia and Vicente, Rodrigo and Blas, Diego and Jenkins, Alexander C. and Barausse, Enrico},
	year         = 2023,
	journal      = {Phys. Rev. D},
	volume       = 107,
	number       = 12,
	pages        = 124033,
	doi          = {10.1103/PhysRevD.107.124033},
	eprint       = {2301.13228},
	archiveprefix = {arXiv},
	primaryclass = {gr-qc}
}

@article{Inchauspe:2024ibs,
	title        = {{Measuring gravitational wave memory with LISA}},
	author       = {Inchausp{\'e}, Henri and Gasparotto, Silvia and Blas, Diego and Heisenberg, Lavinia and Zosso, Jann and Tiwari, Shubhanshu},
	year         = 2025,
	journal      = {Phys. Rev. D},
	volume       = 111,
	number       = 4,
	pages        = {044044},
	doi          = {10.1103/PhysRevD.111.044044},
	eprint       = {2406.09228},
	archiveprefix = {arXiv},
	primaryclass = {gr-qc}
}

@article{Cogez:2026frh,
    author = "Cogez, Adrien and Gasparotto, Silvia and Zosso, Jann and Inchausp{\'e}, Henri and Pitte, Chantal and Maga{\~n}a Zertuche, Lorena and Petiteau, Antoine and Besancon, Marc",
    title = "{Detectability of gravitational-wave memory with LISA: A Bayesian approach}",
    eprint = "2601.23230",
    archivePrefix = "arXiv",
    primaryClass = "gr-qc",
    doi = "10.1103/b9ld-dqq4",
    journal = "Phys. Rev. D",
    volume = "113",
    number = "10",
    pages = "104034",
    year = "2026"
}

@article{Islam:2021old,
	title        = {{Survey of gravitational wave memory in intermediate mass ratio binaries}},
	author       = {Islam, Tousif and Field, Scott E. and Khanna, Gaurav and Warburton, Niels},
	year         = 2023,
	journal      = {Phys. Rev. D},
	volume       = 108,
	number       = 2,
	pages        = {024046},
	doi          = {10.1103/PhysRevD.108.024046},
	eprint       = {2109.00754},
	archiveprefix = {arXiv},
	primaryclass = {gr-qc}
}

@article{Sennett:2016klh,
	title        = {{Gravitational waveforms in scalar-tensor gravity at 2PN relative order}},
	author       = {Sennett, Noah and Marsat, Sylvain and Buonanno, Alessandra},
	year         = 2016,
	journal      = {Phys. Rev. D},
	volume       = 94,
	number       = 8,
	pages        = {084003},
	doi          = {10.1103/PhysRevD.94.084003},
	eprint       = {1607.01420},
	archiveprefix = {arXiv},
	primaryclass = {gr-qc}
}

@article{Bernard:2022noq,
	title        = {{Gravitational waves in scalar-tensor theory to one-and-a-half post-Newtonian order}},
	author       = {Bernard, Laura and Blanchet, Luc and Trestini, David},
	year         = 2022,
	journal      = {JCAP},
	volume       = {08},
	number       = {08},
	pages        = {008},
	doi          = {10.1088/1475-7516/2022/08/008},
	eprint       = {2201.10924},
	archiveprefix = {arXiv},
	primaryclass = {gr-qc}
}

@article{Trestini:2024zpi,
	title        = {{Quasi-Keplerian parametrization for eccentric compact binaries in scalar-tensor theories at second post-Newtonian order and applications}},
	author       = {Trestini, David},
	year         = 2024,
	journal      = {Phys. Rev. D},
	volume       = 109,
	number       = 10,
	pages        = 104003,
	doi          = {10.1103/PhysRevD.109.104003},
	eprint       = {2401.06844},
	archiveprefix = {arXiv},
	primaryclass = {gr-qc}
}

@article{Antoniou:2021zoy,
    author = "Antoniou, Georgios and Leh{\'e}bel, Antoine and Ventagli, Giulia and Sotiriou, Thomas P.",
    title = "{Black hole scalarization with Gauss-Bonnet and Ricci scalar couplings}",
    eprint = "2105.04479",
    archivePrefix = "arXiv",
    primaryClass = "gr-qc",
    doi = "10.1103/PhysRevD.104.044002",
    journal = "Phys. Rev. D",
    volume = "104",
    number = "4",
    pages = "044002",
    year = "2021"
}

@article{LIGOScientific:2026qni,
    author = "Abac, A. G. and others",
    collaboration = "LIGO Scientific, VIRGO, KAGRA",
    title = "{GWTC-4.0: Tests of General Relativity. I. Overview and General Tests}",
    eprint = "2603.19019",
    archivePrefix = "arXiv",
    primaryClass = "gr-qc",
    reportNumber = "LIGO-P2500065",
    month = "3",
    year = "2026"
}

@article{LIGOScientific:2021sio,
    author = "Abbott, R. and others",
    collaboration = "LIGO Scientific, VIRGO, KAGRA",
    title = "{Tests of General Relativity with GWTC-3}",
    eprint = "2112.06861",
    archivePrefix = "arXiv",
    primaryClass = "gr-qc",
    reportNumber = "LIGO-P2100275",
    doi = "10.1103/PhysRevD.112.084080",
    journal = "Phys. Rev. D",
    volume = "112",
    number = "8",
    pages = "084080",
    year = "2025"
}

@article{Ma:2024bed,
    author = "Ma, Sizheng and Nelli, Kyle C. and Moxon, Jordan and Scheel, Mark A. and Deppe, Nils and Kidder, Lawrence E. and Throwe, William and Vu, Nils L.",
    title = "{Einstein{\textendash}Klein{\textendash}Gordon system via Cauchy-characteristic evolution: computation of memory and ringdown tail}",
    eprint = "2409.06141",
    archivePrefix = "arXiv",
    primaryClass = "gr-qc",
    doi = "10.1088/1361-6382/adaf6f",
    journal = "Class. Quant. Grav.",
    volume = "42",
    number = "5",
    pages = "055006",
    year = "2025"
}

@article{Lang:2014osa,
    author = "Lang, Ryan N.",
    title = "{Compact binary systems in scalar-tensor gravity. III. Scalar waves and energy flux}",
    eprint = "1411.3073",
    archivePrefix = "arXiv",
    primaryClass = "gr-qc",
    doi = "10.1103/PhysRevD.91.084027",
    journal = "Phys. Rev. D",
    volume = "91",
    number = "8",
    pages = "084027",
    year = "2015"
}

@article{East:2021bqk,
    author = "East, William E. and Ripley, Justin L.",
    title = "{Dynamics of Spontaneous Black Hole Scalarization and Mergers in Einstein-Scalar-Gauss-Bonnet Gravity}",
    eprint = "2105.08571",
    archivePrefix = "arXiv",
    primaryClass = "gr-qc",
    doi = "10.1103/PhysRevLett.127.101102",
    journal = "Phys. Rev. Lett.",
    volume = "127",
    number = "10",
    pages = "101102",
    year = "2021"
}

@article{Doneva:2021tvn,
    author = "Doneva, Daniela D. and Yazadjiev, Stoytcho S.",
    title = "{Beyond the spontaneous scalarization: New fully nonlinear mechanism for the formation of scalarized black holes and its dynamical development}",
    eprint = "2107.01738",
    archivePrefix = "arXiv",
    primaryClass = "gr-qc",
    doi = "10.1103/PhysRevD.105.L041502",
    journal = "Phys. Rev. D",
    volume = "105",
    number = "4",
    pages = "L041502",
    year = "2022"
}

\clearpage

\end{document}